\documentclass[12pt]{article}
\usepackage{latexsym}
\usepackage{amsfonts}

\newcommand{\mysection}[1]{\section{#1}\setcounter{equation}{0}}

\newcommand{\bea}{\begin{eqnarray}} 
\newcommand{\eea}{\end{eqnarray}}
\newcommand{\beann}{\begin{eqnarray*}} 
\newcommand{\eeann}{\end{eqnarray*}}
\newcommand{\beq}{\begin{equation}} 
\newcommand{\eeq}{\end{equation}}
\newcommand{\ba}{\begin{array}} 
\newcommand{\ea}{\end{array}}
\newcommand{\ben}{\begin{enumerate}} 
\newcommand{\een}{\end{enumerate}}

\newcommand{\5}{\bar }  
\newcommand{\6}{\partial }

\newcommand{\9}[2]{{#1}{}^{{(#2)}}}

\newcommand{\sfrac}[2]{\mbox{$\frac{{#1}}{{#2}}$}\,}

\newcommand{\Ii}{{\mathrm{i}}}

\newcommand{\ep}{\epsilon}
\newcommand{\vep}{\varepsilon}

\newcommand{\da}{{\dot\alpha}} 

\begin{document}

\thispagestyle{empty}

\begin{flushright}
hep-th/0005086
\\
ITP-UH-07/00
\end{flushright}

\vspace{.5cm}

\begin{center}
{\Large
New N=2 supersymmetric gauge theories:\\[4pt]
The double tensor multiplet and its interactions
}
\end{center}
\vspace{.2cm}

\begin{center}
{\large
Friedemann Brandt
}
\end{center}
\vspace{.2cm}

{\sl 
\begin{center} 
Institut f\"ur Theoretische Physik, Universit\"at Hannover,\\
Appelstr.\ 2,
D-30167 Hannover, Germany\\
E-mail: brandt@itp.uni-hannover.de
\end{center}
}
\vspace{.2cm}

\begin{abstract}
The double tensor multiplet of D=4, N=2 supersymmetry, relevant
to type IIB superstring vacua, is
derived and its gauge invariant and N=2 supersymmetric 
interactions are analysed, both self-interactions and interactions
with vector multiplets and hypermultiplets.
Using deformation theory, it is shown that the lowest dimensional 
nontrivial interaction vertices of this type have dimension 5 and
all dimension 5 vertices are determined. They give
rise to new N=2 supersymmetric gauge theories of the
`exotic' type which are local but nonpolynomial in some of the
fields and coupling constants. Explicit examples of such models
are constructed.
\end{abstract}

PACS numbers: 11.30.Pb, 11.15.-q, 11.25.Mj

Keywords: extended supersymmetry, N=2 double tensor multiplet,
new supersymmetric gauge theories

\newpage
\setcounter{page}{1}

\mysection{Introduction and summary}

The paper is devoted to a special multiplet of
four dimensional N=2 supersymmetry
containing two real scalar fields, two 2-form gauge potentials
and two Weyl fermions. This multiplet is particularly relevant
to four dimensional type IIB superstring vacua with N=2 supersymmetry 
because there the
dilaton is a member of such a multiplet, see e.g.\ \cite{LF}. 
The multiplet has been termed double tensor multiplet
in the literature. However, to my knowledge,
it has not been constructed
explicitly up to know. This might have to do 
with the failure of an off-shell construction.
In fact, one may be tempted to expect that
the multiplet, with the above-mentioned field content,
exists off-shell since the number of bosonic and
fermionic degrees of freedom balance off-shell. However, as
shown in section \ref{TT}, such an
off-shell realization of the N=2 supersymmetry 
algebra (possibly with a central charge, and modulo gauge transformations)
is not compatible with 
the free action for the minimal field content.

Of course,
this does not exclude that an off-shell formulation
with additional (auxiliary) fields may exist but
this question will not be
addressed here. We shall thus work with the minimal field content
and derive in section \ref{TT} the N=2 supersymmetry transformations
for the free action.
The N=2 supersymmetry 
algebra is realized on-shell modulo gauge transformations 
of the 2-form gauge potentials, without a central charge.
The gauge transformations that occur in the algebra depend
explicitly and at most linearly on the spacetime coordinates.%
\footnote{The occurrence of gauge transformations involving explicitly
the spacetime coordinates can be understood from a duality
relation between the double tensor multiplet and the hypermultiplet.
This will be explained in more detail and generality in a separate
work.} 
In terms of N=1 multiplets, the double tensor multiplet
consists of two linear multiplets.

Having derived the free double tensor multiplet, we then
study its interactions. Apart from self-interactions, the
interactions with N=2 vector multiplets and hypermultiplets
are analysed because of their
relevance in the string theory context.%
\footnote{The interactions with the N=2 supergravity
multiplet are not discussed here.}
We impose that the action of the interacting theory
be Poincar\'e invariant, gauge invariant and N=2 supersymmetric,
but with possibly modified gauge and supersymmetry transformations
as compared to the free theory.
More precisely, we will study Poincar\'e invariant
nontrivial continuous deformations of the
free theory for one N=2 double tensor and an arbitrary number 
of vector multiplets and hypermultiplets. The analysis uses
a systematic approach which is based on
an expansion in the deformation
parameters (coupling constants) and is briefly reviewed
in section \ref{DEF}. 
The first order deformations must be nontrivial 
on-shell invariants of the free theory, i.e., field polynomials
which are invariant on-shell (modulo total derivatives) under the
gauge and supersymmetry transformations of the free theory.

These on-shell invariants can be determined
at each mass dimension separately (assigning the standard
dimensions to the fields).
In section \ref{DEFS} all such invariants
with dimensions $\leq 5$ are determined.
It turns out that there are no such invariants
with dimensions $\leq 4$; 
in particular, there are thus no nontrivial
power counting renormalizable couplings
of one double tensor multiplet to vector or hypermultiplets at all.
All nontrivial couplings with dimension 5 are trilinear in the
fields. There are three different types of such couplings:

Type A: Self-couplings of the double tensor multiplet. They
contain Freedman-Townsend interaction vertices \cite{FT}.

Type B: Couplings of the double tensor multiplet to vector multiplets.
These couplings are linear in the fields of the 
double tensor multiplet and in the fields of 
different vector multiplets; hence, couplings of this type
involve at least two vector multiplets. They contain
interaction vertices of the Henneaux-Knaepen type \cite{HK}.

Type C: Couplings of the double tensor multiplet to hypermultiplets.
These couplings are linear in the fields of the double tensor multiplet
and quadratic in the fields of one or more hypermultiplets.

In section \ref{HIGHER} we study whether
these dimension 5 interaction vertices
can be extended to higher orders 
in the deformation parameter(s). 
To that end we first
reformulate the free double tensor multiplet
by introducing an appropriate
set of auxiliary fields.
Then we point out a remarkable property common to
all three types of first order deformations
and discuss its consequences for the
structure of the supersymmetry
deformations. Finally we treat explicitly three examples,
one for each coupling type described above.

The first example is the simplest one and 
arises from a type C coupling between the double tensor
multiplet and one hypermultiplet. We complete this
coupling to all orders in the deformation parameters.
In this case the gauge transformations
do not get deformed. In contrast, the supersymmetry
transformations get deformed, without changing the supersymmetry algebra
(the commutator of two deformed supersymmetry transformations
is an ordinary translation plus a gauge transformation on-shell).
The auxiliary fields mentioned above allow one to give
the deformed action and supersymmetry transformations
in a compact polynomial form.
Upon elimination of the
auxiliary fields, the deformed action and the supersymmetry
transformations
become non-polynomial in the deformation parameters
and in the scalar fields of the hypermultiplet, 
but remain local (in fact, 
each term in the
action is at most quadratic in derivatives of fields and
the supersymmetry transformations of the
elementary fields contain at most one derivative).

The second example arises from a coupling of type B between
the double tensor
multiplet and two vector multiplets.
It is somewhat reminiscent of
the N=2 supersymmetric gauge theories
with the vector-tensor multiplet constructed in \cite{VT,T} where
the central charge of that multiplet was gauged
(even though there is no central charge in the present case).
In this case both the supersymmetry
transformations and the gauge transformations
get nontrivially deformed. Again, the supersymmetry
algebra does not change: the commutator of two deformed
supersymmetry transformations is a translation plus a deformed
gauge transformation on-shell.
As in the first example, the complete deformations of the
action, gauge and supersymmetry transformations
are given in a compact polynomial form
using the auxiliary fields. 
Upon elimination of the
auxiliary fields, the deformations of the action and
symmetry transformations become non-polynomial 
in the deformation parameters and in the
scalar and vector fields of the vector multiplets, 
but remain local.

As a third example we discuss
the self-interactions of type A.
Their completion to all orders is
more involved and is not fully accomplished.
However, the first order deformations of the
gauge and supersymmetry transformations and
the first and second order deformation
of the action are computed explicitly in the formulation
with the auxiliary fields.
The result implies already that, 
in the formulation without auxiliary fields, the action
and symmetry transformations
would be non-polynomial in the deformation parameters and
in the scalar fields and the 2-form gauge potentials.
Furthermore it strongly suggests
that all higher order deformations exist as well and
allows one to guess the structure of the resulting 
full action and symmetry 
transformations.

Of course it should be stressed that these three examples
are relatively simple N=2 supersymmetric gauge theories involving
the double tensor multiplet. 
There may be more complicated models of this type. In particular,
instead of discussing the couplings of type A, B or C separately,
one may study linear combinations of them and investigate
whether such linear combinations can be completed to higher orders.
In fact, N=1 supersymmetric models of this more complicated type have
been constructed in \cite{BT} and this suggests that
analogous N=2 supersymmetric models exist as well. 
Section \ref{superfields} comments
on the use of the N=1 superfield construction in \cite{BT}
in the present context. 
One may go even further and study whether couplings
of type A, B or C can be combined with the well-known
couplings relating only vector multiplets and hypermultiplets.
Of course, there are many more open questions, such as
the classification of first order interaction terms
with dimensions $\geq 6$, and the coupling of the
double tensor multiplet to N=2 supergravity.

\mysection{The free double tensor multiplet}
\label{TT}

\paragraph{General ansatz.}
The starting point is the standard
free action for the above-described field content,
\beq
\int d^4x\, 
(\6_\mu a^i \6^\mu a^i-H_\mu^i H^{\mu i}-\Ii \chi\6\5\chi
-\Ii \psi\6\5\psi)
\label{TT1}
\eeq
where the $a^i$ are the two real scalar fields ($i=1,2$), 
$\psi$ and $\chi$ are
the two 2-component Weyl fermions 
(their complex conjugates are denoted by
$\5\psi$ and $\5\chi$), 
and $H_\mu^i$ are the Hodge-duals of the field
strengths of the real 2-form gauge potentials $B_{\mu\nu}^i$,
\beq
H^{\mu i}=\sfrac 12 \ep^{\mu\nu\rho\sigma}\6_\nu B_{\rho\sigma}^i \ .
\label{TT2}
\eeq
Furthermore, here and throughout the paper,
summation over repeated indices of any kind (whether up or down) is
understood (indices $i$ are never lowered).%
\footnote{The remaining
conventions and notation are analogous to those in
\cite{WB}, except that the Minkowski metric
$\eta_{\mu\nu}=\mathit{diag}(1,-1,-1,-1)$ is used.}

To examine whether this free action is N=2 supersymmetric and to determine
the supersymmetry transformations, we make
the most general ansatz for linear N=2 supersymmetry
transformations
compatible with Poincar\'e covariance and
with the dimensions of the fields.
We write these transformations in the form
\beq
\9 \delta0_\xi=\xi^{\alpha i} \9 D0_\alpha{}^i
+\5\xi^i_\da \9 {\5D}0{}^{\da i}\ .
\label{TT3}
\eeq
Here $\xi^i_\alpha$ are two constant anticommuting Weyl spinors
(they are the parameters of the supersymmetry transformations), 
$\9 D0_\alpha{}^i$ are the generators of the corresponding
supersymmetry transformations and
$\5\xi^i_\da$ and $\9 {\5D}0_\da{}^i$ are the complex conjugates
of $\xi^i_\alpha$ and $\9 D0_\alpha{}^i$ respectively.
The ansatz takes then the form
\bea
\9 D0_\alpha{}^i a^j&=&
\sfrac 12 (M^{ij}\chi_\alpha+N^{ij}\psi_\alpha)
\nonumber\\
\9 D0_\alpha{}^i B_{\mu\nu}^j&=&
R^{ij}(\sigma_{\mu\nu}\chi)_\alpha+S^{ij}(\sigma_{\mu\nu}\psi)_\alpha
\nonumber\\
\9 D0_\alpha{}^i\chi_\beta&=&0
\nonumber\\
\9 D0_\alpha{}^i\psi_\beta&=&0
\nonumber\\
\9 D0_\alpha{}^i\5\chi_\da&=&
X^{ij}\6_{\alpha\da}a^j+Y^{ij}H_{\alpha\da}^j
+Z^{ij}\sigma^\mu_{\alpha\da}\6^\nu B_{\nu\mu}^j
\nonumber\\
\9 D0_\alpha{}^i\5\psi_\da&=&
\7X^{ij}\6_{\alpha\da}a^j+\7Y^{ij}H_{\alpha\da}^j
+\7Z^{ij}\sigma^\mu_{\alpha\da}\6^\nu B_{\nu\mu}^j
\label{TT9}
\eea
where the coefficients
$M^{ij}$, \dots , $\7Z^{ij}$ are complex numbers which
are to be determined
from the following requirements: (i) the free Lagrangian must
be invariant modulo a total derivative
under the above transformations, for any choice of
$\xi^i_\alpha$; 
(ii) the commutators of these 
transformations must
fulfill the N=2 supersymmetry algebra at least on-shell, 
modulo gauge transformations and possibly with some central charges.
Clearly, if there is a set of coefficients
$M^{ij}$, \dots , $\7Z^{ij}$ which fulfills these requirements,
it cannot be unique because the Lagrangian is also
invariant under separate $SO(2)$ transformations of
the $a^i$ and $B_{\mu\nu}^i$ and under $SU(2)$ and
$U(1)$ transformations of
the fermions.
Hence, there is some freedom in writing the
supersymmetry transformations owing to these global symmetries. 

(i) is equivalent to the requirement that all
$\9 D0_\alpha{}^i$-transformations
of the free Lagrangian be total derivatives (the
$\9 {\5D}0_\da{}^i$-transformations do not give additional
conditions since
the free Lagrangian is real modulo a total
derivative).
It is straightforward to verify that this imposes
precisely the following conditions:
\beq
\Ii X^{ij}=M^{ij},\quad  \Ii \7X^{ij}=N^{ij},\quad
Y^{ij}=-R^{ij},\quad  \7Y^{ij}=-S^{ij},\quad
Z^{ij}=0=\7Z^{ij}.
\eeq
Using this, the ansatz reduces to
\bea
\9 D0_\alpha{}^i a^j&=&
\sfrac 12 (M^{ij}\chi_\alpha+N^{ij}\psi_\alpha)
\nonumber\\
\9 D0_\alpha{}^i B_{\mu\nu}^j&=&
R^{ij}(\sigma_{\mu\nu}\chi)_\alpha+S^{ij}(\sigma_{\mu\nu}\psi)_\alpha
\nonumber\\
\9 D0_\alpha{}^i\chi_\beta&=&0
\nonumber\\
\9 D0_\alpha{}^i\psi_\beta&=&0
\nonumber\\
\9 D0_\alpha{}^i\5\chi_\da&=&
-\Ii M^{ij}\6_{\alpha\da}a^j-R^{ij}H_{\alpha\da}^j
\nonumber\\
\9 D0_\alpha{}^i\5\psi_\da&=&
-\Ii N^{ij}\6_{\alpha\da}a^j-S^{ij}H_{\alpha\da}^j\ .
\label{TT9a}
\eea
The remaining coefficients are subject to requirement (ii).

\paragraph{Absence of an off-shell representation.}
There are no 
coefficients $M^{ij}$, $N^{ij}$, $R^{ij}$,
$S^{ij}$ such that the transformations 
(\ref{TT9a}) give an off-shell representation
of the standard N=2 supersymmetry algebra (modulo gauge transformations 
and possibly with a central charge). Such an off-shell 
representation would require in particular
$\{\9 D0_\alpha{}^i,\9 {\5D}0_\da{}^j\}=-\Ii\, \delta^{ij}\6_{\alpha\da}$
on the fermions. However, this leads to inconsistent 
equations for the coefficients $M^{ij}$, $N^{ij}$, $R^{ij}$, $S^{ij}$.
Indeed, one finds
\beann
&
\{\9 D0_\alpha{}^i,\9 {\5D}0_\da{}^j\}\chi_\beta=
-\Ii\, \delta^{ij}\6_{\alpha\da}\chi_\beta\ ,\
\{\9 D0_\alpha{}^i,\9 {\5D}0_\da{}^j\}\psi_\beta=
-\Ii\, \delta^{ij}\6_{\alpha\da}\psi_\beta
&
\\
&
\Leftrightarrow\quad
M M^\dagger=
N N^\dagger=
R R^\dagger=
S S^\dagger=1,
\quad
M N^\dagger=
R S^\dagger=0
&
\eeann
where $M$, $N$, $R$, $S$ are the matrices with entries
$M^{ij}$, $N^{ij}$, $R^{ij}$, $S^{ij}$ respectively.

\paragraph{Determination of the supersymmetry transformations.}
Coefficients $M^{ij}$, $N^{ij}$, $R^{ij}$, $S^{ij}$ which
yield an on-shell representation of the N=2 supersymmetry algebra
can be found
by dualizing two scalar fields of an N=2 hypermultiplet.
A hypermultiplet
contains two complex scalar fields $\varphi^i$ ($i=1,2$)
and two Weyl fermions (see section \ref{DEFS} for details).
One may
decompose the scalar fields into real and imaginary parts,
$\varphi^i=a^i+\Ii b^i$, and then ``dualize'' the imaginary parts
according to $\6_\mu b^i\rightarrow \ep^{ji}H_\mu^j$
(the use of $\ep^{ji}$
is a pure convention; other choices are available owing to the
above-mentioned
freedom in writing the supersymmetry transformations).
The transformations 
of the $B_{\mu\nu}^i$ are chosen such that the transformations
of $\6_\mu b^i$ and $\ep^{ji}H_\mu^j$ coincide on-shell, i.e.,
when the free field equations
for the fermions are used.
This dualization procedure yields automatically transformations
that fulfill on-shell the N=2 supersymmetry algebra modulo
gauge transformations.
Furthermore the resulting transformations are indeed of the 
form (\ref{TT9a}).
Hence, they are the sought N=2 supersymmetry transformations
for the double tensor multiplet.
One finds
\bea
\9 D0_\alpha{}^i a^j&=&
\sfrac 12 (\delta^{ij}\chi_\alpha+\ep^{ij}\psi_\alpha)
\nonumber\\
\9 D0_\alpha{}^i B_{\mu\nu}^j&=&
\ep^{ij}(\sigma_{\mu\nu}\chi)_\alpha+\delta^{ij}(\sigma_{\mu\nu}\psi)_\alpha
\nonumber\\
\9 D0_\alpha{}^i\chi_\beta&=&0
\nonumber\\
\9 D0_\alpha{}^i\psi_\beta&=&0
\nonumber\\
\9 D0_\alpha{}^i\5\chi_\da&=&
-\Ii\, \6_{\alpha\da}a^i-\ep^{ij}H_{\alpha\da}^j
\nonumber\\
\9 D0_\alpha{}^i\5\psi_\da&=&
-\Ii\, \ep^{ij}\6_{\alpha\da}a^j-H_{\alpha\da}^i\ .
\label{TT9c}
\eea
{}From these transformations one reads off that
the double tensor multiplet consists of two
N=1 linear multiplets. For instance, the 
N=1 linear multiplets with respect to $\9 D0_\alpha{}^1$
are $(a^1,B_{\mu\nu}^2,\chi)$ and
$(a^2,B_{\mu\nu}^1,\psi)$.

\paragraph{N=2 supersymmetry algebra.}
As remarked above, the transformations (\ref{TT9c})
yield an on-shell representation of the N=2 supersymmetry algebra
because they can be obtained 
by dualizing two scalar fields of
a hypermultiplet. One finds that the
algebra is realized on-shell
modulo gauge transformations of the $B_{\mu\nu}^i$, 
without a central charge.
More precisely, the algebra is realized even off-shell on the
scalar fields,
\bea
\{\9 D0_\alpha{}^i,\9 {\5D}0_\da{}^j\}a^k&=&
-\Ii\, \delta^{ij}\6_{\alpha\da}a^k
\nonumber\\
\{\9 D0_\alpha{}^i,\9 D0_\beta{}^j\}a^k&=&0
\label{TT10}
\eea
while it holds on-shell on the fermions,
\bea
\{\9 D0_\alpha{}^i,\9 {\5D}0_\da{}^j\}\chi_\beta&=&
-\Ii\, \delta^{ij}\6_{\alpha\da}\chi_\beta
-\Ii\, \ep^{ij}\ep_{\alpha\beta} \6_{\gamma\da}\psi^\gamma
\approx -\Ii\, \delta^{ij}\6_{\alpha\da}\chi_\beta 
\nonumber\\
\{\9 D0_\alpha{}^i,\9 D0_\beta{}^j\}\chi_\gamma &=&0
\nonumber\\
\{\9 D0_\alpha{}^i,\9 D0_\beta{}^j\}\5\chi_\da &=&
\Ii\, \ep^{ij}\ep_{\alpha\beta} \6_{\gamma\da}\psi^\gamma
\approx 0
\nonumber\\
\{\9 D0_\alpha{}^i,\9 {\5D}0_\da{}^j\}\psi_\beta&=&
-\Ii\, \delta^{ij}\6_{\alpha\da}\psi_\beta
+\Ii\, \ep^{ij}\ep_{\alpha\beta} \6_{\gamma\da}\chi^\gamma
\approx -\Ii\, \delta^{ij}\6_{\alpha\da}\psi_\beta
\nonumber\\
\{\9 D0_\alpha{}^i,\9 D0_\beta{}^j\}\psi_\gamma &=&0
\nonumber\\
\{\9 D0_\alpha{}^i,\9 D0_\beta{}^j\}\5\psi_\da &=&
-\Ii\, \ep^{ij}\ep_{\alpha\beta} \6_{\gamma\da}\chi^\gamma
\approx 0
\label{TT11}
\eea
where $\approx$ denotes on-shell equality.
Finally, on the 2-form gauge potentials one has
\bea
\{\9 D0_\alpha{}^i,\9 {\5D}0_\da{}^j\}B_{\mu\nu}^k&=&
-\Ii\, \delta^{ij}\6_{\alpha\da}B_{\mu\nu}^k
+2\6^{}_{[\mu} \Lambda^{ijk}_{\nu]\alpha\da}
-2\ep^{ij}\ep^{kl}x_{\alpha\da} \6^{}_{[\mu} H_{\nu]}^l
\nonumber\\
&\approx&
-\Ii\, \delta^{ij}\6_{\alpha\da}B_{\mu\nu}^k
+\6_\mu \Lambda^{ijk}_{\nu\alpha\da}
-\6_\nu \Lambda^{ijk}_{\mu\alpha\da}
\nonumber\\
\{\9 D0_\alpha{}^i,\9 D0_\beta{}^j\}B_{\mu\nu}^k&=&0
\label{TT12}
\eea
where
\beq
\Lambda^{ijk}_{\mu\nu}=
\sfrac{\Ii}{2} \eta_{\mu \nu}(\delta^{jk}\ep^{il}+
\delta^{il}\ep^{jk}+\delta^{ik}\ep^{jl}+\delta^{jl}\ep^{ik})a^l
+\ep^{ij}\ep^{kl}x_\nu H_\mu^l-\Ii\, \delta^{ij} B_{\mu\nu}^k\ .
\label{TT13}
\eeq
Note that the terms with the $\Lambda^{ijk}_{\mu\nu}$ in 
(\ref{TT12}) are special gauge transformations.
Hence, the N=2 supersymmetry algebra is indeed represented on
the 2-form gauge potentials on-shell modulo gauge
transformations with ``gauge parameters''
involving the $\Lambda^{ijk}_{\mu\nu}$.

To summarize,
in the notation (\ref{TT3}) one has on all fields
\beq
[\9 \delta0_\xi,\9 \delta0_{\xi'}]\approx
\xi^\mu \6_\mu + \9 \delta0_\Lambda
\eeq
where $\xi^\mu$ is a constant vector involving the supersymmetry
parameters,
\beq
\xi^\mu=\Ii\, (\xi^{i\prime}\sigma^\mu\5\xi^i
-\xi^i\sigma^\mu\5\xi^{i\prime}),
\label{TT14}
\eeq
and $\9 \delta0_\Lambda$ is a special gauge transformation,
\beq
\9 \delta0_\Lambda B_{\mu\nu}^i=\6_\mu \Lambda_\nu^i-\6_\nu \Lambda_\mu^i\ ,
\quad \Lambda_\mu^i=\Lambda^{jki}_{\mu\nu}\,
(\xi^j\sigma^\nu\5\xi^{k\prime}-\xi^{j\prime}\sigma^\nu\5\xi^k).
\label{TT15}
\eeq
Note that both these gauge transformations and the terms with equations
of motion which appear in the commutator on $B_{\mu\nu}^i$ involve 
explicitly the spacetime coordinates $x^\mu$.
It is easy to check that the sum of these terms actually does not depend
explicitly on the $x^\mu$ (it cannot because
there is no explicit $x$-dependence in the supersymmetry transformations):
the relevant terms in the first anticommutator (\ref{TT12}) combine to
the $x$-independent term
$2\ep^{ij}\ep^{kl}H^l_{[\nu}\sigma_{\mu]\alpha\da}$.
However, this term is not to be interpreted as some
kind of ``central charge'' in the supersymmetry algebra because
it equals a gauge transformation on-shell. The point is that this
gauge transformation depends explicitly on the $x^\mu$, owing to
\beann
H^l_{[\nu}\sigma_{\mu]\alpha\da}
=
\6^{}_{[\mu}(H^l_{\nu]}x_{\alpha\da} )-x_{\alpha\da} \6^{}_{[\mu} H^l_{\nu]}
\approx \6^{}_{[\mu}(H^l_{\nu]}x_{\alpha\da} ).
\eeann
A more complete discussion of the r\^ole and origin of
such terms in supersymmetry algebras will be given elsewhere.

\mysection{Brief description of deformation theory}
\label{DEF}

To study interactions involving the double tensor multiplet, we shall
start from the free action for one double tensor and
an arbitrary number of vector multiplets and
hypermultiplets. We shall seek
deformations of the free action which are invariant under the standard
Poincar\'e transformations and under possibly deformed gauge and
supersymmetry transformations. The requirement that these
deformations be continuous means that the deformed
Lagrangian $L$, the corresponding gauge transformations $\delta_\vep$ and 
supersymmetry transformations $\delta_\xi$ can be expanded in
deformation parameters (coupling constants) according to
\bea
L&=&\9 L0 + \9 L1 + \9 L2 + \dots
\label{i0}\\
\delta_\vep&=&\9 \delta0_\vep + \9 \delta1_\vep + \9 \delta2_\vep + \dots
\label{i2}\\
\delta_\xi&=&\9 \delta0_\xi + \9 \delta1_\xi + \9 \delta2_\xi + \dots
\label{i3}\\
\9 \delta{k}_\xi&=&\xi^{\alpha i}\9 Dk_\alpha{}^i
+\5\xi^i_\da \9 {\5D}k{}^{\da i}
\label{i3a}
\eea
where $\9 Lk$, $\9 \delta{k}_\vep$, $\9 \delta{k}_\xi$,
$\9 Dk_\alpha{}^i$ and $\9 {\5D}k{}^{\da i}$ have
order $k$ in the deformation parameters, and
$\9 L0$, $\9 \delta0_\vep$ and $\9 \delta0_\xi$ denote the free Lagrangian,
its gauge symmetries and supersymmetry transformations respectively.
$\vep$ and $\xi$ denote collectively the
``parameters'' of
gauge and N=2 supersymmetry transformations respectively, 
i.e., the $\vep$'s are
arbitrary fields whereas the $\xi$'s are constant
anticommuting spinors.

The invariance requirements in the deformed theory are
\beq
\delta_\vep L\simeq 0,\quad \delta_\xi L\simeq 0,
\label{i1}
\eeq
where $\simeq$ is equality modulo total derivatives.
The analysis can now be performed
``perturbatively'' by expanding these invariance requirements
in the deformation parameters.
At first order, this requires
\bea
&& \9 \delta0_\vep \9 L1+\9 \delta1_\vep \9 L0\simeq 0
\label{i4}\\
&& \9 \delta0_\xi \9 L1+\9 \delta1_\xi \9 L0\simeq 0.
\label{i5}\eea
These equations can also be cast in the form
\bea
&& \9 \delta0_\vep \9 L1+\sum_\Phi (\9 \delta1_\vep \Phi)
\, \frac{\delta \9 L0}{\delta \Phi}\simeq 0
\label{i4a}\\
&& \9 \delta0_\xi \9 L1+\sum_\Phi (\9 \delta1_\xi \Phi)
\, \frac{\delta \9 L0}{\delta \Phi}\simeq 0
\label{i5a}\eea
where the sum $\sum_\Phi$ runs over all fields
and $\delta \9 L0/\delta \Phi$ is the
Euler-Lagrange derivative of the free Lagrangian
with respect to $\Phi$. This
shows that $\9 L1$ has to be invariant on-shell under the
zeroth order transformations $\9 \delta0_\vep$ and
$\9 \delta0_\xi$, where this on-shell invariance
refers to the free field equations $\delta \9 L0/\delta \Phi=0$.
Furthermore, we are only interested in nontrivial deformations, i.e.\ in
deformations that cannot be removed through mere field redefinitions.
The free Lagrangian
changes under infinitesimal field redefinitions $\Delta\Phi$
through terms $\sum_\Phi (\Delta\Phi)\delta \9 L0/\delta \Phi
+\6_\mu M^\mu$. These are terms which vanish on-shell in the free
theory modulo total derivatives. Terms of this form are thus
trivial and can be neglected without loss of generality. 

Hence, the first step of the perturbative approach to
the deformation problem is the determination of nontrivial
on-shell invariants of the free theory. 
The first order deformation of the
free Lagrangian is a linear combination of these on-shell invariants.
The corresponding first order deformations of the gauge 
and supersymmetry transformations are the coefficient functions of
the Euler-Lagrange derivatives
$\delta \9 L0/\delta \Phi$ which appear in (\ref{i4a}) and (\ref{i5a}).

At second order, (\ref{i1}) gives
\bea
&& \9 \delta0_\vep \9 L2+\9 \delta1_\vep \9 L1+
\sum_\Phi (\9 \delta2_\vep \Phi)
\, \frac{\delta \9 L0}{\delta \Phi}\simeq 0
\label{i6}\\
&& \9 \delta0_\xi \9 L2+\9 \delta1_\xi \9 L1+
\sum_\Phi (\9 \delta2_\xi \Phi)
\, \frac{\delta \9 L0}{\delta \Phi}\simeq 0.
\label{i7}\eea
These equations require that
$\9 \delta1_\vep \9 L1$ and $\9 \delta1_\xi \9 L1$ be in the
image of $\9 \delta0_\vep$ and $\9 \delta0_\xi$ respectively,
at least on-shell (with respect to the free theory)
and modulo total derivatives. This
can impose relations between the coefficients
of the on-shell invariants in $\9 L1$ (it can even set some of
these coefficients to zero). Additional relations between
these coefficients can be imposed by the equations
arising from (\ref{i1}) at even higher orders of the deformation problem.

Such relations between the coefficients in $\9 L1$ have
a cohomological characterization. In fact, the whole deformation theory
sketched above can be usefully reformulated as a cohomological problem
in the framework of an extended BRST formalism. 
The relevant cohomology is that of an extended BRST differential $\9 s0$
which encodes the global supersymmetry transformations,
the gauge transformations and the equations of motion of the free theory.
This cohomological formulation of the deformation theory
is described in \cite{BHWFB} and extends the
deformation theory developed in \cite{BH}.
In the cohomological approach, the classification
of the on-shell invariants of the free theory amounts to compute
the cohomology of the extended BRST differential of the free
theory  in the space of local functionals with ghost number 0.
The deformation problem at orders $\geq 2$ is
controlled by the same cohomology, but now at ghost number 1:
the cohomology classes at ghost number 1 give the 
possible obstructions
to the existence of a deformation at orders $\geq 2$.
This r\^ole of the cohomology at ghost number 1
is similar to the characterization of
candidate anomalies through the BRST cohomology in the
quantum field theoretical context.

\mysection{First order deformations of dimension $\leq 5$}
\label{DEFS}

\subsection{Free action}

The input for the deformation theory is
the free action for one double tensor multiplet and
a set of vector multiplets and hypermultiplets.%
\footnote{The generalization to
the case with more
than one double tensor multiplet is left to the reader.}
The fields of a vector multiplet are a real gauge field $A_\mu$,
a complex scalar field $\phi$ and two Weyl fermions $\lambda^i_\alpha$.
The complex conjugates of $\phi$ and $\lambda^i_\alpha$ are denoted by
$\5\phi$ and $\5\lambda^i_\da$ respectively (note: as before, complex
conjugation does not
lower the index $i$).
The fields of a hypermultiplet are two complex scalar fields
$\varphi^i$ and two Weyl fermions $\rho_\alpha$
and $\eta_\alpha$. Their complex conjugates are
denoted by $\5\varphi^i$, $\5\rho_\da$ and $\5\eta_\da$. 
The free Lagrangian is 
\bea
\9 L0&=&
\6_\mu a^i \6^\mu a^i-H_\mu^i H^{\mu i}-\Ii \chi\6\5\chi
-\Ii \psi\6\5\psi
\nonumber\\
&&
-\sfrac 14 F^A_{\mu\nu}F^{\mu\nu A}
+\sfrac 12 \6_\mu\phi^A\6^\mu \5\phi^A
-2\Ii \lambda^{iA}\6 \5\lambda^{iA}
\nonumber\\
&&
+\6_\mu\varphi^{ia}\6^\mu \5\varphi^{ia}-\Ii \rho^a\6 \5\rho^a
-\Ii \eta^a\6 \5\eta^a
\eea
where $A$ and $a$ label the vector multiplets and 
hypermultiplets respectively and $F^A_{\mu\nu}$ is the field strength of
$A_\mu^A$,
\[
F^A_{\mu\nu}=\6_\mu A_\nu^A-\6_\nu A_\mu^A\ .
\]
Using again the notation (\ref{TT3}), the
N=2 supersymmetry transformations of the vector multiplets and hypermultiplets
read
\bea
\9 D0_\alpha{}^i A^A_\mu &=& \ep^{ij}(\sigma_\mu \5\lambda^{jA})_\alpha
\nonumber\\
\9 D0_\alpha{}^i \phi^A &=& 2 \lambda^{iA}_\alpha
\nonumber\\
\9 D0_\alpha{}^i \5\phi^A &=& 0 
\nonumber\\
\9 D0_\alpha{}^i \lambda^{jA}_\beta &=& -\sfrac{\Ii}2 \ep^{ij}
                    \sigma^{\mu\nu}_{\alpha\beta}F^A_{\mu\nu}
\nonumber\\
\9 D0_\alpha{}^i \5\lambda^{jA}_\da &=& -\sfrac {\Ii}2 \delta^{ij}\, 
\6_{\alpha\da}\5\phi^A
\nonumber\\
\9 D0_\alpha{}^i \varphi^{ja} &=& \ep^{ij}\rho^a_\alpha
\nonumber\\
\9 D0_\alpha{}^i \5\varphi^{ja} &=& \delta^{ij}\eta^a_\alpha
\nonumber\\
\9 D0_\alpha{}^i \rho^a_\beta &=& 0
\nonumber\\
\9 D0_\alpha{}^i \eta^a_\beta &=& 0
\nonumber\\
\9 D0_\alpha{}^i \5\rho^a_\da &=& -\Ii\,\ep^{ij}\6_{\alpha\da}\5\varphi^{ja}
\nonumber\\
\9 D0_\alpha{}^i \5\eta^a_\da &=& -\Ii\,\6_{\alpha\da}\varphi^{ia}
\label{D1}
\eea
The gauge symmetries act nontrivially only on $A^A_\mu$ and $B_{\mu\nu}^i$
according to
\bea
\9 {\delta_\vep}0 A^A_\mu &=& \6_\mu\vep^A
\\
\9 {\delta_\vep}0 B_{\mu\nu}^i &=& \6_\mu \vep^i_\nu - \6_\nu \vep^i_\mu\ .
\label{D2}
\eea

\subsection{General remarks and strategy}

As described in the previous section, the first step
within the deformation approach is the determination
of the nontrivial on-shell invariants of the free theory.
The classification of these on-shell invariants can be
carried out separately in subspaces of field polynomials
with definite
dimension and degree in the fields of the various multiplets.

To specify these subspaces,
we assign dimension 1 to all bosons (scalar fields, vector fields,
2-form gauge potentials) and to the spacetime derivatives $\6_\mu$, 
dimension 3/2 to all
fermions, and dimension 0 to the gauge parameters
$\vep^A$ and $\vep^i_\mu$. 
Then the supersymmetry
transformations $\9 D0_\alpha{}^i$ have dimension 1/2 and
the gauge transformations
$\9 {\delta_\vep}0$ have dimension 0.
Furthermore the $\9 D0_\alpha{}^i$
are linear in the fields
and do not mix fields of different supermultiplets.
One can therefore classify
the on-shell invariants separately in subspaces
of field polynomials characterized by $(d,N_{TT},N_V,N_H)$
where $d$ is the dimension, and 
$N_{TT}$, $N_V$ and $N_H$ are the degrees in the fields
of the double tensor multiplet, vector multiplets and hyper
multiplets respectively.

To specify the field polynomials in a given subspace, it is helpful
to denote by $N_b$, $N_f$ and $N_\6$ the degree in
the bosons, fermions and spacetime derivatives respectively.
A field polynomial with a definite degree $N_\Phi$ in all fields
and a definite dimension $d$ fulfills thus
\beq
N_b+N_f=N_\Phi,\quad N_b+N_\6+\sfrac 32 N_f=d.
\label{D3}
\eeq
This yields in particular
\beq
N_\6+\sfrac 12 N_f=d-N_\Phi.
\label{D4}
\eeq
A field polynomial characterized by $(d,N_{TT},N_V,N_H)$
has $N_\Phi=N_{TT}+N_V+N_H$ and thus
can contain only terms with $(N_\6,N_f)=(d-N_\Phi,0)$, $(d-N_\Phi-1,2)$,
\dots, $(0,2d-2N_\Phi)$.

Note that $N_\Phi$ ranges from
1 to $d$ (for given value of $d$). It is easy to verify that the values 
$N_\Phi=1$ and $N_\Phi=d$ do not give
nontrivial on-shell invariants because we impose also 
Poincar\'e invariance.
Indeed, in both cases the only Poincar\'e invariant
field polynomials which are nontrivial and on-shell
gauge invariant modulo a total derivative are polynomials
in the undifferentiated scalar fields
(there are gauge invariants with
$N_\Phi=1$, such as each $A^A_\mu$, but they are not Poincar\'e invariant); 
no such polynomial is
on-shell supersymmetric modulo a total derivative.
In particular, there are no on-shell invariants with $d=2$.
In the following we shall discuss the remaining cases with
$d=3,4,5$ and $N_{TT}\geq 1$.
The discussion covers both true interaction terms
and terms quadratic in the fields.

The computations have been
done in two steps:
1. Determination of
the most general nontrivial real Poincar\'e invariant
field polynomial $P_{(d,N_{TT},N_V,N_H)}$
which satisfies $\9 \delta0_\vep P_{(d,N_{TT},N_V,N_H)}\sim 0$
where $\sim$ denotes on-shell equality in the free theory
modulo a total derivative; 2. Imposing
$\9 D0_\alpha{}^i P_{(d,N_{TT},N_V,N_H)}\sim 0$ which
yields then the first order
deformations $\9 L1_{(d,N_{TT},N_V,N_H)}$ for each
case $(d,N_{TT},N_V,N_H)$ separately.
Of course, $P_{(d,N_{TT},N_V,N_H)}$ and
$\9 L1_{(d,N_{TT},N_V,N_H)}$ are determined only
modulo trivial terms which do not matter.
We will give ``minimal expressions'' for these
polynomials containing as many strictly gauge invariant terms
as possible, and as few trivial terms as possible.

\subsection{d=3} 

The polynomials to be discussed are those
with $(d,N_{TT},N_V,N_H)=(3,2,0,0)$, $(3,1,0,1)$
and $(3,1,1,0)$ respectively. The only Poincar\'e invariant terms 
$(d,N_{TT},N_V,N_H)=(3,2,0,0)$ 
are bilinears in the fermions 
of the double tensor multiplet, $P_{(3,2,0,0)}=
k_1\5\chi\5\chi+k_2\5\psi\5\psi+k_3\5\chi\5\psi+\mathrm{c.c.}$ where 
$k_1,k_2,k_3\in\mathbb{C}$. It is straightforward to verify that
no such term is on-shell supersymmetric
modulo a total derivative (unless $k_1=k_2=k_3=0$) because the
$H_\mu^i$ occur in the transformation of $\5\psi$ and $\5\chi$.

Similarly, the only Poincar\'e invariant terms
with $(d,N_{TT},N_V,N_H)=(3,1,0,1)$
are linear combinations of bilinears in fermions, with one fermion of the
double tensor multiplet and one fermion of a hypermultiplet
respectively. Again, no nonvanishing linear combination of this
type is on-shell supersymmetric
modulo a total derivative owing to the presence of the
$H_\mu^i$ in the transformations of $\5\psi$ and $\5\chi$.

The case $(d,N_{TT},N_V,N_H)=(3,1,1,0)$ is more involved
because now there are both bilinears in the fermions and terms
with bosons which are Poincar\'e invariant and on-shell gauge invariant
modulo a total derivative,
\[
P_{(3,1,1,0)}=k_1^i A_\mu \6^\mu a^i+
k_2^i A_\mu H^{\mu i}+k_3^i \lambda^i\psi+k_4^i \lambda^i\chi
+\5k_3^i \5\lambda^i\5\psi+\5k_4^i \5\lambda^i\5\chi
\]
where $k_1^i,k_2^i\in\mathbb{R}$, $k_3^i,k_4^i\in\mathbb{C}$
(since the supersymmetry transformations do not
mix the fields of different vector multiplets, the
discussion can be made for each vector
multiplet separately and we have dropped the index $A$).
Modulo trivial terms,
$\9 D0_\alpha{}^iP_{(3,1,1,0)}$ contains
terms proportional to $F_{\mu\nu}\sigma^{\mu\nu}\psi$,
$F_{\mu\nu}\sigma^{\mu\nu}\chi$ and $H^j_{\alpha\da}\5\lambda^{\da k}$.
$\9 D0_\alpha{}^iP_{(3,1,1,0)}\sim 0$
requires that the coefficients of these terms vanish. This imposes
\beann
&\ep^{ij}k_1^j-\Ii k_2^i+\Ii \ep^{ij}k_3^j=0&
\\
&k_1^i-\Ii\ep^{ij} k_2^j+\Ii \ep^{ij}k_4^j=0&
\\
&\ep^{ik}k_2^j-\delta^{ij}\5 k_3^k-\ep^{ij}\5 k_4^k=0&
\eeann
which yield
\[
k_1^i=k_2^i=k_3^i=k_4^i=0.
\]

{\em Remark.} Recall that the cases
$N_{TT}=0$ are not discussed here. In fact, there are
nontrivial on-shell invariants with $d=3$ and $N_{TT}=0$.
These are the mass terms for the fermions of
the hypermultiplets which have $(d,N_{TT},N_V,N_H)=(3,0,0,2)$.
As the above discussion shows, these mass terms have no counterparts
with $N_{TT}>0$ owing to the presence of the 2-form gauge potentials
in the double tensor multiplet (more precisely: the
presence of
$H_\mu^i$ in the supersymmetry transformations of 
$\5\psi$ and $\5\chi$).

\subsection{d=4}

We need to discuss the cases $N_\Phi=2$ and $N_\Phi=3$.
The cases $N_\Phi=2$ are easy: there are simply no nontrivial terms
which are quadratic in the fields,
Poincar\'e invariant and on-shell gauge invariant
modulo a total derivative. Indeed, the candidate terms
$(d,N_{TT},N_V,N_H)=(4,2,0,0)$ are linear combinations
of the terms $H_\mu^i H^{\mu j}$, $H_\mu^i \6^\mu a^j$,
$\6_\mu a^i \6^\mu a^j$, $\psi\6\5\psi$, $\chi\6\5\chi$,
$\chi\6\5\psi$ (modulo trivial ones); 
candidate terms
$(d,N_{TT},N_V,N_H)=(4,1,1,0)$ or $(4,1,0,1)$ are
linear combinations of terms such as
$\6_\mu a^i \6^\mu \phi^A$, $H_\mu^i \6^\mu \varphi^a$,
$\lambda^{iA}\6\5\psi$ etc.;
all these terms vanish on-shell (in the free theory) 
modulo a total derivative.
Note that the free Lagrangian itself is of this type
and clearly vanishes 
on-shell modulo a total derivative.

The various terms with $N_\Phi=3$ and $N_{TT}\geq 1$
are $(d,N_{TT},N_V,N_H)=(4,3,0,0)$,
$(4,2,1,0)$, $(4,2,0,1)$, $(4,1,2,0)$, $(4,1,1,1)$, $(4,1,0,2)$.
Owing to (\ref{D4}), the field polynomials in these
subsectors contain only
terms with $(N_\6,N_f)=(1,0)$ or $(0,2)$.

Consider first the case $(d,N_{TT},N_V,N_H)=(4,3,0,0)$. 
Poincar\'e invariance excludes all terms with $(N_\6,N_f)=(1,0)$.
The most general Poincar\'e invariant
field polynomial with $(N_\6,N_f)=(0,2)$ is
$k_1^i a^i\5\psi\5\psi+k_2^i a^i\5\psi\5\chi
+k_3^i a^i\5\chi\5\chi+k_4^i B_{\mu\nu}^i\5\psi\5\sigma^{\mu\nu}\5\chi
+\mathrm{c.c.}$ The requirement that it be gauge invariant
on-shell modulo a total derivative yields $k_4^i=0$, i.e.,
$P_{(4,3,0,0)}=k_1^i a^i\5\psi\5\psi+k_2^i a^i\5\psi\5\chi
+k_3^i a^i\5\chi\5\chi+\mathrm{c.c.}$.
$\9 D0_\alpha{}^i P_{(4,3,0,0)}\sim 0$
imposes $k_1^i=k_2^i=k_3^i=0$.
 
An analogous discussion shows that the
other cases with $N_V=0$ [$(d,N_{TT},N_V,N_H)=(4,2,0,1)$ 
and $(4,1,0,2)$]
do not yield nontrivial on-shell invariants.

The remaining cases $(d,N_{TT},N_V,N_H)=(4,2,1,0)$, 
$(4,1,2,0)$ and $(4,1,1,1)$
are more involved because now there are terms
with $(N_\6,N_f)=(1,0)$ and a new type of terms with
$(N_\6,N_f)=(0,2)$ which are Poincar\'e invariant and
gauge invariant on-shell modulo total derivatives.
These terms are of the type $A^A_\mu j_A^\mu$ where
$j_A^\mu$ are Noether currents of the free theory which have
dimension 3 and are bilinear in the fields.

In the case $(d,N_{TT},N_V,N_H)=(4,2,1,0)$
we can drop the index $A$ labelling
the vector multiplets owing to $N_V=1$ (each vector multiplet
can be treated separately because the supersymmetry transformations do not
mix the fields of different vector multiplets). 
The general form of $j^\mu$ in $A_\mu j^\mu$ is in this case
\[
j^\mu=k_1 \ep^{ij} a^i\6^\mu a^j+k_2 \psi\sigma^\mu\5\psi
+k_3 \chi\sigma^\mu\5\chi
+k_4 \psi\sigma^\mu\5\chi+\5k_4 \chi\sigma^\mu\5\psi
\]
where $k_1,k_2,k_3\in\mathbb{R}$, $k_4\in\mathbb{C}$
(this $j^\mu$ is thus  a linear combination of five
different real Noether currents).
In addition, there are terms of the type met already above,
involving one scalar field and
two fermions (and no derivative),
\[
\phi (k_5\5\psi\5\psi+k_6\5\chi\5\psi+k_7\5\chi\5\chi)
+\5\phi (k_8\5\psi\5\psi+k_9\5\chi\5\psi+k_{10}\5\chi\5\chi)
+ a^i\5\lambda^j(k_{11}^{ij}\5\psi+k_{12}^{ij}\5\chi)+\mathrm{c.c.}
\]
with $k_5,\dots,k_{12}^{ij}\in\mathbb{C}$.
It is easy to verify that
all coefficients $k_5,\dots,k_{12}^{ij}$ must vanish. This comes again
from the fact that the $\9 D0_\alpha{}^i$-transformations
of $\5\chi$ and $\5\psi$ contain the $H_\mu^i$; as a consequence,
$\9 D0_\alpha{}^i(k_5\phi\5\psi\5\psi+\dots)$
contains in particular terms with one
scalar field and one $H$, i.e.\ terms $\phi H \5\psi$,
$\5\phi H \5\psi$, $\phi H \5\chi$, $\5\phi H \5\chi$,
$a H\5\lambda$.
Such terms do not occur in $\9 D0_\alpha{}^i(A_\mu j^\mu)$.
This enforces
$k_5=\dots=k_{12}^{ij}=0$.
Finally,
$\9 D0_\alpha{}^i(A_\mu j^\mu)\sim 0$
imposes $k_1=k_2=k_3=k_4=0$. This simply follows from the
fact that $\9 D0_\alpha{}^i(A_\mu j^\mu)$ involves terms
with $\5\lambda$'s owing to
$\9 D0_\alpha{}^iA_\mu=\ep^{ij}(\sigma_\mu \5\lambda^{j})_\alpha$;
evidently $\ep^{ij}(\sigma_\mu \5\lambda^{j})_\alpha j^\mu\sim 0$ imposes
$k_1=k_2=k_3=k_4=0$.

The remaining cases $(d,N_{TT},N_V,N_H)=(4,1,2,0)$ and $(4,1,1,1)$
are similar. Again there terms of the form $A^A_\mu j_A^\mu$
with $j$'s of the form $s\6s+f\5f$ and terms
$s\5f\5f+\mathrm{c.c.}$ where $s$ and $f$ stand
for scalar fields and fermions respectively. The coefficients
of all terms $s\5f\5f$ containing $\5\psi$ or $\5\chi$ must
vanish because of the presence of $H$ in the supersymmetry
transformations of $\5\psi$ or $\5\chi$. The remaining
terms $s\5f\5f+\mathrm{c.c.}$ have necessarily $s=a^1$ or $s=a^2$ 
and do not contain fermions of the double tensor multiplet
(due to $N_{TT}=1$); their coefficients must vanish because
their supersymmetry
transformations contain 3-fermion-terms
$(\9 D0_\alpha{}^ia^j)ff$ which do not occur in 
$\9 D0_\alpha{}^i(A_\mu j^\mu)$.
It is then easy to verify that the terms $A^A_\mu j_A^\mu$
must vanish as well.

{\em Remark.} Again, the case $N_{TT}=0$ yields nontrivial
on-shell invariants: 
$(d,N_{TT},N_V,N_H)=(4,0,3,0)$ gives the N=2 supersymmetric
extension of the cubic vertices between the vector gauge fields, 
$(d,N_{TT},N_V,N_H)=(4,0,1,2)$ gives couplings between vector
and hypermultiplets of the form $A^A_\mu j_A^\mu+\dots$\ .
These on-shell invariants yield of course the deformations
of the free Lagrangian for vector multiplets and hypermultiplets 
to the standard
$N=2$-supersymmetric abelian or nonabelian gauge theories.
 
\subsection{d=5}

We shall first discuss the polynomials with $N_\Phi=3$
as they yield nontrivial
first order deformations. The cases to be discussed are 
$(d,N_{TT},N_V,N_H)=(5,3,0,0)$, $(5,1,2,0)$, $(5,1,0,2)$,
$(5,2,1,0)$, $(5,2,0,1)$, $(5,1,1,1)$. The first three
cases yield the interaction vertices of type A, B and C
mentioned in the introduction.
It is helpful
to observe that, for $d=5$ and $N_\Phi=3$, there are only
terms with $(N_\6,N_f)=(2,0)$ and $(1,2)$, owing to (\ref{D4}).
Furthermore, there are no nontrivial $(N_\6,N_f)=(2,0)$-terms
involving three scalar fields and no nontrivial
$(N_\6,N_f)=(1,2)$-terms involving a scalar field.
Indeed, every Poincar\'e invariant $(N_\6,N_f)=(2,0)$-term
with three scalar fields $s_1,s_2,s_3$ can be brought
to the form $s_1\6_\mu s_2\6^\mu s_3$ by adding trivial terms.
However, $s_1\6_\mu s_2\6^\mu s_3$ itself is trivial because
in the free theory it is on-shell equal
to $\frac 12\6_\mu(s_1s_2\6^\mu s_3-s_3s_2\6^\mu s_1+s_1s_3\6^\mu s_2)$.
Every Poincar\'e invariant $(N_\6,N_f)=(1,2)$-term with
a scalar field $s$ is 
modulo a total derivative a linear
combination of terms $sf\6\5f$ and $s(\6f)\5f$ which 
vanish on-shell in the free theory. This makes it relatively
easy to determine $P_{(d,N_{TT},N_V,N_H)}$ in the various cases.
I note that in order to compute $\9 D0_\alpha{}^i P_{(d,N_{TT},N_V,N_H)}$, 
it is often useful to employ
\[
\9 D0_\alpha{}^i H_\mu^j
\approx-\sfrac{\Ii}{2} (\ep^{ij}\6_\mu \chi_\alpha
+\delta^{ij}\6_\mu \psi_\alpha)
\]
where $\approx$ is again on-shell equality in the free theory.

\paragraph{First order interactions of type A.}
We start with the case $(d,N_{TT},N_V,N_H)=(5,3,0,0)$
and treat it in some detail to illustrate the
calculations. In this case one finds (modulo trivial terms)
\beann
P_{(5,3,0,0)}&=&
\sfrac 12 k_1^k \ep^{ij}H_\mu^iH_\nu^jB_{\rho\sigma}^k\ep^{\mu\nu\rho\sigma}
+k_2^{ijk}H_\mu^iH^{\mu j}a^k+k_3^k\ep^{ji} a^i\6^\mu a^j H_\mu^k
\\
&&+(k_4^i\chi\sigma^\mu\5\chi+k_5^i\psi\sigma^\mu\5\psi
+k_6^i\psi\sigma^\mu\5\chi+\5k_6^i\chi\sigma^\mu\5\psi)H_\mu^i
\eeann
where $k_1^i,k_2^{ijk},k_3^i,k_4^i,k_5^i\in\mathbb{R}$,
$k_6^i\in\mathbb{C}$ and (without loss of generality)
$k_2^{ijk}=k_2^{jik}$. One now computes
$\9 D0_\alpha{}^i P_{(5,3,0,0)}$. Up to trivial terms the result
is a linear combination of terms
of the form $\psi HH$, $\chi HH$, $\psi H\6a$, $\chi H\6a$.
There are two types of $\psi HH$-terms:
terms $H_\mu^jH_\nu^k\sigma^{\mu\nu}\psi$ which are antisymmetric
in $jk$ (one has $H_\mu^jH_\nu^k\sigma^{\mu\nu}\psi
=\frac 12\ep^{jk}\ep^{lm}H_\mu^lH_\nu^m\sigma^{\mu\nu}\psi$)
and terms $H_\mu^jH^{\mu k}\psi$ which are symmetric in $jk$.
Similarly there are two types of $\chi HH$-terms.
Vanishing of the coefficients of 
$H_\mu^jH_\nu^k\sigma^{\mu\nu}\psi$
and $H_\mu^jH_\nu^k\sigma^{\mu\nu}\chi$ imposes
\[
\Ii k_1^i+\ep^{ij}k_5^j-k^i_6=0,\quad
\Ii k_1^i+\ep^{ij}k_4^j+\5k^i_6=0.
\]
Vanishing of the coefficients of 
$H_\mu^jH^{\mu k}\psi$ and $H_\mu^jH^{\mu k}\chi$ imposes
\beann
-\Ii \ep^{ij}k_1^k+\sfrac 12\ep^{il}k_2^{jkl}+
\delta^{ij}k^k_5+\ep^{ij}k^k_6+(j\leftrightarrow k)=0
\\
\Ii \delta^{ij}k_1^k+\sfrac 12k_2^{jki}+
\ep^{ij}k^k_4+\delta^{ij}\5k^k_6+(j\leftrightarrow k)=0.
\eeann
Vanishing of the coefficients of
$\psi H_\mu^k\6^\mu a^j$ and $\chi H_\mu^k\6^\mu a^j$ imposes:
\beann
-\Ii k_2^{ikj}-\delta^{ij}k^k_3-2\Ii \ep^{ij}k^k_5-2\Ii\delta^{ij}k^k_6=0
\\
-\Ii \ep^{il}k_2^{lkj}+\ep^{ij}k^k_3
-2\Ii \delta^{ij}k^k_4-2\Ii\ep^{ij}\5k^k_6=0.
\eeann
These equations give
\[
k_2^{ijk}=k^i_4=k^i_5=\mathrm{Re}\,k^i_6=0,\quad
k^i_3=2k_1^i,\quad \mathrm{Im}\,k^i_6=k_1^i.
\]
Choosing $k_1^i$ as deformation parameters,
the resulting nontrivial first order deformation
is thus
\bea
&
\9 L1_{(5,3,0,0)}=
\sfrac 12 k_1^k \ep^{ij}H_\mu^iH_\nu^jB_{\rho\sigma}^k\ep^{\mu\nu\rho\sigma}
-2k_1^iH_\mu^i\ep^{jk} a^j\6^\mu a^k 
+\Ii k_1^iH_\mu^i
(\psi\sigma^\mu\5\chi-\chi\sigma^\mu\5\psi),&
\nonumber\\
&k_1^i\in\mathbb{R}\ .
\label{L300}
\eea

\paragraph{First order interactions of type B.}
The case $(d,N_{TT},N_V,N_H)=(5,1,2,0)$ is more complex. One has
\beann
P_{(5,1,2,0)}&=& H_\mu^i A_\nu^A (k_1^{iAB}F^{\mu\nu B}
+k_2^{iAB}F_{\rho\sigma}^B\ep^{\mu\nu\rho\sigma})
+a^i F_{\mu\nu}^A (k_3^{iAB}F^{\mu\nu B}
+k_4^{iAB} F_{\rho\sigma}^B\ep^{\mu\nu\rho\sigma})
\\
&&
+[H_\mu^i \phi^A \6^\mu (k_5^{iAB}\5\phi^B
+k_6^{iAB}\phi^B)
+k_7^{ijAkB}H_\mu^i\lambda^{jA}\sigma^\mu\5\lambda^{kB}
\\
&&
+F_{\mu\nu}^A\lambda^{iB}\sigma^{\mu\nu}(k_8^{AiB}\chi
+k_9^{AiB}\psi)
+\mathrm{c.c.}]
\eeann
where $k_1^{iAB},\dots,k_4^{iAB}\in\mathbb{R}$,
$k_5^{iAB},\dots,k_9^{AiB}\in\mathbb{C}$, and (without loss
of generality)
$k_2^{iAB}=k_2^{iBA}$, $k_3^{iAB}=k_3^{iBA}$, $k_4^{iAB}=k_4^{iBA}$,
$k_5^{iAB}=-\5k_5^{iBA}$, $k_6^{iAB}=-k_6^{iBA}$, 
$k_7^{ijAkB}=\5k_7^{ikBjA}$
(e.g., $k_6^{iAB}=-k_6^{iBA}$ can be imposed owing to
$H_\mu^i\phi^{(A}\6^\mu\phi^{B)}=\sfrac 12 H_\mu^i\6^\mu
(\phi^{A}\phi^{B})\simeq 0$).
$\9 D0_\alpha{}^i P_{(5,1,2,0)}\sim 0$ imposes
\beann
& k_2^{iAB}=k_3^{iAB}=k_4^{iAB}=\mathrm{Im}\,k_5^{iAB}=
k_6^{iAB}=k_8^{AiB}=k_9^{AiB}=
\mathrm{Re}\,k_7^{ijAkB}=0,&
\\
& 
k_1^{iAB}=-k_1^{iBA}=-2\mathrm{Re}\,k_5^{iAB},\quad
\mathrm{Im}\,k_7^{ijAkB}=-k_1^{iAB}\delta^{jk}.
\eeann
This gives 
\bea
&\9 L1_{(5,1,2,0)}= k_1^{iAB}H_\mu^i
(A_\nu^A F^{\mu\nu B}
-\sfrac 12 \phi^A \6^\mu \5\phi^B-\sfrac 12\5\phi^A \6^\mu \phi^B
-2\Ii\lambda^{jA}\sigma^\mu\5\lambda^{jB}),&
\nonumber\\
& k_1^{iAB}=-k_1^{iBA}\in\mathbb{R}\ .&
\label{L120}
\eea

\paragraph{First order interactions of type C.}
The case $(d,N_{TT},N_V,N_H)=(5,1,0,2)$ is quite simple. One has
\beann
P_{(5,1,0,2)}&=& H_\mu^i(k_1^{ijakb}\varphi^{ja}\6^\mu\varphi^{kb}
+k_2^{ijakb}\varphi^{ja}\6^\mu\5\varphi^{kb}
\\
&&
+k_3^{iab}\rho^a\sigma^\mu\5\rho^b+k_4^{iab}\eta^a\sigma^\mu\5\eta^b
+k_5^{iab}\rho^a\sigma^\mu\5\eta^b)+\mathrm{c.c.}
\eeann
where $k_1^{ijakb},\dots,k_5^{iab}\in\mathbb{C}$,
and, without loss of generality,
$k_1^{ijakb}=-k_1^{ikbja}$, $k_2^{ijakb}=-\5k_2^{ikbja}$,
$k_3^{iab}=\5k_3^{iba}$, $k_4^{iab}=\5k_4^{iba}$.
$\9 D0_\alpha{}^i P_{(5,1,0,2)}\sim 0$ imposes
in addition $k_3^{iab}=-\5k_4^{iab}$, 
$k_2^{ijakb}=-2\Ii \delta^{jk}k_3^{iab}$ and
$k_1^{ijakb}=\Ii\ep^{jk}k_5^{iab}$.
This yields
\bea
&\9 L1_{(5,1,0,2)}=
k_3^{iab} H_\mu^i (-2\Ii \varphi^{ja}\6^\mu\5\varphi^{jb}
+\rho^a\sigma^\mu\5\rho^b-\eta^b\sigma^\mu\5\eta^a)&
\nonumber\\
&
\phantom{\9 L1_{(5,1,0,2)}=}
+k_5^{iab} H_\mu^i
(\Ii\ep^{jk}\varphi^{ja}\6^\mu\varphi^{kb}+\rho^a\sigma^\mu\5\eta^b)
+\mathrm{c.c.}\ ,&
\nonumber\\
& 
k_3^{iab}=\5k_3^{iba}\in\mathbb{C},\quad
k_5^{iab}=k_5^{iba}\in\mathbb{C}.&
\label{L102}
\eea

\paragraph{Remaining cases.}
The remaining cases do not give nontrivial first order deformations
and will therefore be discussed only briefly.
In the case
$(d,N_{TT},N_V,N_H)=(5,2,1,0)$ one has
\beann
P_{(5,2,1,0)}=\phi H^i_\mu (k_1^{ij} H^{\mu j}
+k_2^{ij} \6^\mu a^j)
+k_3 F_{\mu\nu}\chi\sigma^{\mu\nu}\psi
+H_\mu^i \lambda^j\sigma^\mu(k_4^{ij} \5\psi+k_5^{ij} \5\chi)+\mathrm{c.c.}
\eeann
where $k_1^{ij},\dots,k_5^{ij} \in\mathbb{C}$.
$\9 D0_\alpha{}^iP_{(5,2,1,0)}\sim 0$ gives
\[
k_1^{ij}=\dots=k_5^{ij}=0.
\]

The case $(d,N_{TT},N_V,N_H)=(5,1,1,1)$ is similar to the
case $(5,2,1,0)$, owing to the duality
relation between the double tensor multiplet and a hypermultiplet.
As a consequence, it does not
give nontrivial first order deformations.

In the case
$(d,N_{TT},N_V,N_H)=(5,2,0,1)$ one has
\beann
P_{(5,2,0,1)}&=&H^i_\mu
(k_1^{ijk} H^{\mu j}\varphi^k
+k_2^{ijk}\varphi^j\6^\mu a^k+k_3^i\chi\sigma^\mu\5\rho
\\
&&
+k_4^i\psi\sigma^\mu\5\rho
+k_5^i\chi\sigma^\mu\5\eta+k_6^i\psi\sigma^\mu\5\eta)+\mathrm{c.c.}
\eeann
where $k_1^{ijk},\dots,k_6^{i} \in\mathbb{C}$.
Again, $\9 D0_\alpha{}^iP_{(5,2,0,1)}\sim 0$ imposes
\[
k_1^{ijk}=\dots=k_6^{i}=0.
\]

The remaining cases with $d=5$ are those with 
$N_\Phi=2$ and $N_\Phi=4$. The field polynomials with $N_\Phi=2$
contain terms with $(N_\6,N_f)=(3,0)$ and $(2,2)$.
It is easy to see that all Poincar\'e invariant
terms of these types which are gauge invariant
on-shell modulo
total derivatives are trivial (an example is $F^{\mu\nu}\6_\mu H_\nu^i$).
Field polynomials with $N_\Phi=4$ contain only terms 
with $(N_\6,N_f)=(1,0)$ and $(0,2)$;
there are no such Poincar\'e invariant polynomials with $N_{TT}\geq 1$
which are gauge invariant and N=2
supersymmetric modulo trivial terms.

\mysection{New N=2 supersymmetric gauge theories}
\label{HIGHER}

\subsection{Reformulation of the free double tensor multiplet}

The study and construction 
of the deformations at higher orders is considerably simplified
by switching to an alternative (equivalent) formulation of the free
double tensor multiplet.
The free action (\ref{TT1}) 
for the double tensor multiplet is replaced by
\beq
\9 S0_{TT}=
\int d^4x\, 
(\6_\mu a^i \6^\mu a^i+h_\mu^i h^{\mu i}
+2h_\mu^i H^{\mu i}-\Ii \chi\6\5\chi
-\Ii \psi\6\5\psi)
\label{TTfree}
\eeq
where the $h_\mu^i$ are auxiliary vector fields.
Eliminating them by their algebraic equations of motion
reproduces (\ref{TT1}). 
Since (\ref{TT1}) and (\ref{TTfree}) agree except for
the term $(h_\mu^i+H_\mu^i)(h^{\mu i}+H^{\mu i})$,
(\ref{TTfree}) is invariant under the supersymmetry transformations
(\ref{TT9c}) supplemented by 
$\9 D0_\alpha{}^i h_\mu^j=-\9 D0_\alpha{}^i H_\mu^j$.
Furthermore it is convenient to substitute $h$'s for
$H$'s in the supersymmetry transformations of the fermions.
This is achieved
by adding suitable on-shell trivial symmetries to the transformations
of the fermions and the $h$'s. The resulting new supersymmetry
transformations $\9 {\4D}0_\alpha{}^i$ are
\beann
\9 {\4D}0_\alpha{}^i\5\chi_\da&=&
\9 D0_\alpha{}^i\5\chi_\da+\frac{1}{2}\,\sigma^{\mu}_{\alpha\da}\,
\ep^{ij}\,\frac{\delta \9 {L}0}{\delta h^{\mu j}}
\\
\9 {\4D}0_\alpha{}^i\5\psi_\da&=&
\9 D0_\alpha{}^i\5\psi_\da+\frac{1}{2}\,\sigma^{\mu}_{\alpha\da}\,
\delta^{ij}\,\frac{\delta \9 {L}0}{\delta h^{\mu j}}
\\
\9 {\4D}0_\alpha{}^i h^{\mu j}&=&
\9 D0_\alpha{}^i h^{\mu j}-\frac{1}{2}\,\sigma^{\mu}_{\alpha\da}\,
\Big[
\ep^{ij}\,\frac{\delta \9 {L}0}{\delta \5\chi_\da}
+\delta^{ij}\,\frac{\delta \9 {L}0}{\delta \5\psi_\da}
\Big]
\eeann
where $\delta \9 {L}0/\delta \Phi$ is the Euler-Lagrange derivative
of the free Lagrangian in (\ref{TTfree})
with respect to $\Phi$. The new transformations
differ from the supersymmetry transformations (\ref{TT9c}) only
through terms that vanish on-shell. Hence, they fulfill the 
N=2 supersymmetry algebra (on-shell and modulo gauge transformations).
We shall from now on drop the tilde-symbol and
denote the new supersymmetry transformations again by
$\9 D0_\alpha{}^i$,
\bea
\9 D0_\alpha{}^i a^j&=&
\sfrac 12 (\delta^{ij}\chi_\alpha+\ep^{ij}\psi_\alpha)
\nonumber\\
\9 D0_\alpha{}^i B_{\mu\nu}^j&=&
\ep^{ij}(\sigma_{\mu\nu}\chi)_\alpha+\delta^{ij}(\sigma_{\mu\nu}\psi)_\alpha
\nonumber\\
\9 D0_\alpha{}^i h_\mu^j&=&
\sfrac{\Ii}2 \6_\mu (\ep^{ij}\chi_\alpha+\delta^{ij}\psi_\alpha)
\nonumber\\
\9 D0_\alpha{}^i\chi_\beta&=&0
\nonumber\\
\9 D0_\alpha{}^i\psi_\beta&=&0
\nonumber\\
\9 D0_\alpha{}^i\5\chi_\da&=&
-\Ii\, \6_{\alpha\da}a^i+\ep^{ij}h_{\alpha\da}^j
\nonumber\\
\9 D0_\alpha{}^i\5\psi_\da&=&
-\Ii\, \ep^{ij}\6_{\alpha\da}a^j+h_{\alpha\da}^i\ .
\label{TT9d}
\eea
The gauge transformations $\9 \delta0_\vep$ do not change; the auxiliary
fields $h_\mu^i$ are invariant under these transformations. 
The free action and supersymmetry transformations
of the vector multiplets and hypermultiplets are not changed.

The computations involve the following steps:

1. The auxiliary fields $h_\mu^i$ substitute for
$(-H_\mu^i)$ in the first order deformations
(\ref{L300}), (\ref{L120}) and (\ref{L102}).
This is possible because
$h_\mu^i$ and $-H_\mu^i$ coincide on-shell
in the free theory.

2. The second step is the determination of the corresponding 
first order deformations $\9 \delta1_\vep$ and
$\9 D1_\alpha{}^i$
of the gauge and supersymmetry
transformations. That amounts to making the
coefficients of the Euler-Lagrange derivatives
in (\ref{i5}) explicit for the respective
first order deformations.

3. One computes $\9 \delta1_\vep \9 L1$ and 
$\9 D1_\alpha{}^i \9 L1$ and
seeks an $\9 L2$ such that (\ref{i6}) and (\ref{i7})
hold. If necessary, one then proceeds analogously to
higher orders.

\subsection{Structure of the supersymmetry deformations}

To carry out these computations and to understand the
structure of the resulting supersymmetry
deformations, the following observation is useful.
The first order deformations
(\ref{L300}), (\ref{L120}) and (\ref{L102}) have
a remarkable property in common:
in the fomulation with the auxiliary fields $h_\mu^i$,
all of them (and therefore all linear combinations of them too) 
are of the form
\beq
\9 L1=h_\mu^i\, j^{\mu i}
\label{d1}
\eeq
where $j^{\mu i}$ are Noether currents of the free theory,
\beq
\6_\mu j^{\mu i}=\sum_{\Phi} (\Delta^i \Phi)\,
\frac{\delta \9 L0}{\delta \Phi}\ .
\label{d2}
\eeq
Here $\Delta^i$ generates the global symmetry of the free Lagrangian
corresponding by Noether's first theorem to $j^{\mu i}$.
Owing to the supersymmetry transformation
of $h_\mu^i$ in (\ref{TT9d}), one has
\beq
\9 D0_\alpha{}^i\9 L1=
\sfrac{\Ii}2 \6_\mu (\ep^{ij}\chi_\alpha+\delta^{ij}\psi_\alpha) j^{\mu j}
+h_\mu^j\, \9 D0_\alpha{}^i j^{\mu j}.
\label{d3}
\eeq
Owing to (\ref{d2}), the first term on the right hand side
is a linear combination of the Euler-Lagrange derivatives
of $\9 L0$ modulo a total derivative,
\beq
\sfrac{\Ii}2 \6_\mu (\ep^{ij}\chi_\alpha+\delta^{ij}\psi_\alpha) j^{\mu j}
\simeq 
-\sum_{\Phi}
\sfrac{\Ii}2 (\ep^{ij}\chi_\alpha+\delta^{ij}\psi_\alpha)
(\Delta^j \Phi)\,
\frac{\delta \9 L0}{\delta \Phi}\ .
\label{d4}
\eeq
Hence, in order that a term of the form (\ref{d1}) gives a 
supersymmetric first order deformation,
the second term on the right hand side of (\ref{d3})
must also be a linear combination of the Euler-Lagrange derivatives
of $\9 L0$ (modulo a total derivative),
\beq
h_\mu^j\, \9 D0_\alpha{}^i j^{\mu j}\simeq
-\sum_{\Phi} (\Delta^i_\alpha \Phi)\,
\frac{\delta \9 L0}{\delta \Phi}\ ,
\label{d5}
\eeq
for some transformations $\Delta^i_\alpha$.
Equations (\ref{d3}) through (\ref{d5}) yield then
first order deformations of the supersymmetry transformations
of the following form:
\beq
\9 D1_\alpha{}^i\Phi=\Delta^i_\alpha\Phi+
\sfrac{\Ii}2 (\ep^{ij}\chi_\alpha+\delta^{ij}\psi_\alpha)\Delta^j\Phi\ .
\label{d6}
\eeq
Note that (\ref{d4}) holds for any Noether current $j^{\mu i}$,
in contrast to (\ref{d5}). The currents which appear in 
(\ref{L300}), (\ref{L120}) and (\ref{L102}) have thus the
special property to fulfill (\ref{d5}).

Even though the supersymmetry transformations do get nontrivially 
deformed, one would not expect that the supersymmetry algebra
gets deformed (otherwise the deformed supersymmetry algebra
would contain global symmetries generated by transformations
which are at least quadratic in the fields -- a very unlikely
possibility). Indeed one finds in all examples to be
discussed in the following that the
supersymmetry algebra in the deformed model is the
standard one, i.e., it has the form
\beq
{}[ \delta_\xi, \delta_{\xi'}]\approx
\xi^\mu \6_\mu + \delta_\mathrm{gauge}
\label{algebra}
\eeq
where $\approx$ is on-shell-equality in the deformed model (note:
in the previous sections $\approx$ denoted on-shell-equality in the 
free model!), $\delta_\xi$ are the deformed supersymmetry transformations,
$\xi^\mu \6_\mu$ is an ordinary translation with parameter
$\xi^\mu$ as in (\ref{TT14}), and
$\delta_\mathrm{gauge}$ is a deformed gauge transformation.

\subsection{First example (type C)}

The simplest examples of the first order deformations
(\ref{L102}) are those involving only one hypermultiplet.
In these examples we can thus drop the index $a$ distinguishing
different hypermultiplets. Furthermore we choose $k_3^i=0$ and
$k_5^i=\Ii g^i$ with real $g^i$.
In the formulation with 
the auxiliary fields $h_\mu^i$, (\ref{L102}) then becomes
\[
\9 L1=
g^i h_\mu^i
(\ep^{jk}\varphi^{j}\6^\mu\varphi^{k}-\Ii \rho\sigma^\mu\5\eta
+\mathrm{c.c.}),\quad g^i\in\mathbb{R}\ .
\]
\paragraph{Sketch of the computation.}
$\9 L1$ is indeed of the form (\ref{d1}), with $j^{\mu i}=g^i j^\mu$
and $j^\mu$ the term in parenthesis.
We have $\9 \delta0_\vep \9 L1=0$, i.e., the gauge transformations do
not get deformed at first order.
In order to determine the first order deformations of the supersymmetry
transformations one computes $\9 D0_\alpha{}^i \9 L1$. The result
is
\beann
\9 D0_\alpha{}^i \9 L1\simeq
\Gamma^i_\alpha \Big[
\ep^{jk}\varphi^j\,\frac{\delta \9 L0}{\delta \5\varphi^k}
+\ep^{jk}\5\varphi^j\,\frac{\delta \9 L0}{\delta \varphi^k}
+\eta_\beta\,\frac{\delta \9 L0}{\delta \rho_\beta}
-\rho_\beta\,\frac{\delta \9 L0}{\delta \eta_\beta}
+\5\eta_\da\,\frac{\delta \9 L0}{\delta \5\rho_\da}
-\5\rho_\da\,\frac{\delta \9 L0}{\delta \5\eta_\da}
\Big]
\\
+\Ii g^j h_{\alpha\da}^j\Big[
\varphi^i\,\frac{\delta \9 L0}{\delta \5\rho_\da}
-\ep^{ik}\5\varphi^k\,\frac{\delta \9 L0}{\delta \5\eta_\da}
\Big]
+\Ii  g^k(\varphi^i\sigma_{\mu\nu}\rho
      -\ep^{ij}\5\varphi^j\sigma_{\mu\nu}\eta)_\alpha
\,\frac{\delta \9 L0}{\delta B_{\mu\nu}^k}
\eeann
where
\beq
\Gamma^i_\alpha=\sfrac {\Ii}2  g^j
(\ep^{ij}\chi_\alpha+\delta^{ij}\psi_\alpha) .
\label{Gamma}
\eeq
Owing to $j^{\mu i}=g^i j^\mu$, we have $\Delta^i=g^i\Delta$,
and (\ref{d6}) reads
\[
\9 D1_\alpha{}^i\Phi= \Delta^i_\alpha\Phi+\Gamma^i_\alpha \Delta\Phi
\]
where $\Delta^i_\alpha$ acts nontrivially only on $\5\rho$, $\5\eta$
and the $B$'s,
\beann
&&
\Delta^i_\alpha \5\rho_\da=-\Ii g^k h_{\alpha\da}^k\,\varphi^i
\\
&&
\Delta^i_\alpha \5\eta_\da=\Ii g^k h_{\alpha\da}^k\,
\ep^{ij}\5\varphi^j
\\
&&
\Delta^i_\alpha B_{\mu\nu}^j=\Ii\, g^j
(\ep^{ik}\5\varphi^k\sigma_{\mu\nu}\eta
-\varphi^i\sigma_{\mu\nu}\rho)_\alpha
\\
&&
\Delta^i_\alpha (\mbox{other fields})=0,
\eeann
and $\Delta$ rotates the fields of the hypermultiplet,
\bea
&\Delta \varphi^i= \ep^{ij}\5\varphi^j,\quad
 \Delta \5\varphi^i= \ep^{ij}\varphi^j
&
\nonumber\\
&\Delta \rho_\alpha=-\eta_\alpha\ ,\quad
\Delta \eta_\alpha=\rho_\alpha\ ,\quad
\Delta \5\rho_\da=-\5\eta_\da\ ,\quad
\Delta \5\eta_\da=\5\rho_\da
&
\nonumber\\
&
\Delta (\mbox{other fields})=0.
&
\label{Delta1}\eea
Next one computes $\9 D1_\alpha{}^i \9 L1$.
One easily verifies $\Delta \9 L1=0$.
Using in addition $\9 D1_\alpha{}^i h_\mu^j=0$ and
$\9 D0_\alpha{}^i (g^jh_\mu^j)=\6_\mu \Gamma^i_\alpha$, it is
straightforward to verify that one gets
\[
\9 D1_\alpha{}^i \9 L1 = -\9 D0_\alpha{}^i(
g^j h^{\mu j} g^k h_\mu^k\varphi^l\5\varphi^l).
\]
\paragraph{Result.}
This completes the construction of the deformation.
Indeed, since the previous equation
does not involve the free field equations,
we can set $\9 D2_\alpha{}^i=0$ and
the term in parenthesis on the right hand side
can be taken as $\9 L2$:
we have $\9 D1_\alpha{}^i \9 L2=0$ and
$\9 \delta0_\vep \9 L2=0$. Hence, one gets a
deformed Lagrangian $L=\9 L0+\9 L1+\9 L2$ which is 
invariant under the gauge transformations $\9 \delta0_\vep$
and the supersymmetry transformations $D_\alpha^i=
\9 D0_\alpha{}^i+\9 D1_\alpha{}^i$.
The result can be written more compactly in terms of an
auxiliary covariant derivative
\beq
\7D_\mu=\6_\mu-g^i h_\mu^i\Delta
\label{cov}
\eeq 
with $\Delta$ as in (\ref{Delta1}).
The deformed Lagrangian for the double tensor multiplet and the hyper
multiplet reads then
\beq
L=\9 L0_{TT}+\7D_\mu\varphi^{i}\7D^\mu \5\varphi^{i}
-\Ii \rho\sigma^\mu \7D_\mu \5\rho
-\Ii \eta\sigma^\mu \7D_\mu \5\eta
\label{L1}
\eeq
with $\9 L0_{TT}$ as in (\ref{TTfree}).
The deformed supersymmetry transformations are
\bea
D_\alpha^i \varphi^{j} &=& \ep^{ij}\rho_\alpha
+\Gamma^i_\alpha \ep^{jk}\5\varphi^k
\nonumber\\
D_\alpha^i \5\varphi^{j} &=& \delta^{ij}\eta_\alpha
+\Gamma^i_\alpha \ep^{jk}\varphi^k
\nonumber\\
D_\alpha^i \rho_\beta &=& -\Gamma^i_\alpha \eta_\beta
\nonumber\\
D_\alpha^i \eta_\beta &=& \Gamma^i_\alpha \rho_\beta
\nonumber\\
D_\alpha^i \5\rho_\da &=& 
-\Ii\,\ep^{ij}\7D_{\alpha\da}\5\varphi^{j}
-\Gamma^i_\alpha\5\eta_\da
\nonumber\\
D_\alpha^i \5\eta_\da &=& -\Ii\,\7D_{\alpha\da}\varphi^{i}
+\Gamma^i_\alpha\5\rho_\da
\nonumber\\
D_\alpha^i B_{\mu\nu}^j&=&
(\ep^{ij}\sigma_{\mu\nu}\chi+\delta^{ij}\sigma_{\mu\nu}\psi
+\Ii g^j
\ep^{ik}\5\varphi^k\sigma_{\mu\nu}\eta
-\Ii g^j\varphi^i\sigma_{\mu\nu}\rho)_\alpha
\nonumber\\
D_\alpha^i &=&\9 D0_\alpha{}^i\ \mbox{on the other fields.}
\eea
It can now be explicitly verified that
(\ref{algebra}) holds in the deformed model.

One may finally eliminate the auxiliary fields $h_\mu^i$
by their algebraic equations of motion. That amounts
to the identification (both in the action and symmetry
transformations)
\beq
h_\mu^i\equiv
-\sfrac 12 K^{ij}
(H_\mu^j+g^j\ep^{kl}\varphi^{k}\6_\mu\varphi^{l}-\Ii g^j\rho\sigma_\mu\5\eta
+\mathrm{c.c.}),\quad
K^{ij}=\delta^{ij}-\frac{g^ig^j\varphi^k\5\varphi^k}
{1+g^mg^m\varphi^n\5\varphi^n}\ . 
\label{h1}
\eeq

\subsection{Second example (type B)}

The simplest first order deformation (\ref{L120}) arises
when only two vector multiplets are involved. 
So, in the following we take $A=1,2$. This gives
$k^{iAB}=-g^i\ep^{AB}$ in (\ref{L120}) and
\[
\9 L1= g^i h_\mu^i\ep^{AB}
(A_\nu^A F^{\mu\nu B}
-\sfrac 12 \phi^A \6^\mu \5\phi^B-\sfrac 12\5\phi^A \6^\mu \phi^B
-2\Ii\lambda^{jA}\sigma^\mu\5\lambda^{jB}),\quad
g^i\in\mathbb{R}\ .
\]
\paragraph{Sketch of the computation.}
Again $\9 L1$ has the form (\ref{d1}), with $j^{\mu i}=g^i j^\mu$.
This time $\9 \delta0_\vep \9 L1$ does not vanish
(even modulo a total derivative), i.e., the
gauge transformations are deformed.
One has
\beann
\9 \delta0_\vep \9 L1\simeq
-\sfrac 14 g^i\ep^{AB}\vep^A \ep_{\mu\nu\rho\sigma}F^{\rho\sigma B}
\,\frac{\delta \9 L0}{\delta B_{\mu\nu}^i}
+g^i h_\mu^i\ep^{AB}\vep^A\,\frac{\delta \9 L0}{\delta A_\mu^B}\ .
\eeann
The first order deformation of the gauge transformations
is therefore
\beann
\9 \delta1_\vep A_\mu^A&=&g^i h_\mu^i\ep^{AB}\vep^B
\\
\9 \delta1_\vep B_{\mu\nu}^i &=& 
\sfrac 14 g^i\vep^A \ep^{AB}\ep_{\mu\nu\rho\sigma}F^{\rho\sigma B},
\quad\9 \delta1_\vep(\mbox{other fields})=0
\eeann
Next one computes $\9 D0_\alpha{}^i \9 L1$. The result
is again of the form
\[
\9 D0_\alpha{}^i \9 L1\simeq -\sum_\Phi
(\9 D1_\alpha{}^i\Phi)\,\frac{\delta \9 L0}{\delta \Phi}\ ,\quad
\9 D1_\alpha{}^i\Phi
=\Delta^i_\alpha\Phi +\Gamma^i_\alpha \Delta\Phi
\]
with $\Gamma^i_\alpha$ as in (\ref{Gamma}), but
now $\Delta^i_\alpha$ is given by
\beann
&&
\Delta^i_\alpha \lambda^{jA}_\beta=-\Ii\ep^{ij}\ep^{AB}
g^k h_\mu^k\, A_\nu^B\,\sigma^{\mu\nu}{}_{\alpha\beta}
\\
&&
\Delta^i_\alpha \5\lambda^{jA}_\da=-\sfrac {\Ii}2 \delta^{ij}\ep^{AB}
g^k h_{\alpha\da}^k\,\5\phi^B
\\
&&
\Delta^i_\alpha B_{\mu\nu}^j=\Ii g^j\ep^{AB}
(\5\phi^A\sigma_{\mu\nu}\lambda^{iB}
+\ep^{ik}A_{[\mu}^A\sigma^{}_{\nu]}\5\lambda^{kB})_\alpha
\\
&&
\Delta^i_\alpha (\mbox{other fields})=0
\eeann
and $\Delta$ rotates the fields of the vector multiplets,
\beq
\Delta X^A= -\ep^{AB}X^B\quad \mbox{for}\quad
X^A\in\{A_\mu^A,\phi^A,\5\phi^A,\lambda^{Ai}_\alpha,\5\lambda^{iA}_\da\},
\quad
\Delta (\mbox{other fields})=0.
\label{Delta2}
\eeq
To determine the second order deformation, one must
compute both $\9 \delta1_\vep \9 L1$ and
$\9 D1_\alpha{}^i \9 L1$.
As in the first example, $\9 L1$ is $\Delta$-invariant which makes 
it to easy to compute $\9 D1_\alpha{}^i \9 L1$. The result is
\beann
\9 \delta1_\vep \9 L1&=&-\9 \delta0_\vep \9 L2
+\sfrac 12 g^ig^j h^{\rho j} A^{\sigma A} \vep^A\ep_{\mu\nu\rho\sigma}\,
\frac{\delta \9 L0}{\delta B_{\mu\nu}^i}
\\
\9 D1_\alpha{}^i \9 L1&=&-\9 D0_\alpha{}^i\9 L2
\eeann
where
\[
\9 L2=\sfrac 12(g^i h^{\mu i} g^j h^{\nu j} A_\mu^A A_\nu^A
-g^i h^{\mu i} g^j h_\mu^j A^{\nu A} A_\nu^A
+g^i h^{\mu i} g^j h_\mu^j \phi^A\5\phi^A).
\]
The second order deformations of the
gauge and supersymmetry transformations are thus
\[
\9 \delta2_\vep B_{\mu\nu}^i =
-\sfrac 12 g^i  \vep^A\ep_{\mu\nu\rho\sigma}
g^j h^{\rho j} A^{\sigma A},\quad
\9 \delta2_\vep(\mbox{other fields})=0,\quad
\9 D2_\alpha{}^i=0.
\]
\paragraph{Result.}
This completes the construction of the deformation.
Indeed, one has $\9 D1_\alpha{}^i \9 L2=0$ and
$\9 \delta2_\vep \9 L2=0$. 
Hence, $L=\9 L0+\9 L1+\9 L2$ is invariant (modulo total derivatives)
under $D_\alpha^i=\9 D0_\alpha{}^i+\9 D1_\alpha{}^i$
and $\delta_\vep=\9 \delta0_\vep+\9 \delta1_\vep+\9 \delta2_\vep$.
Again, the deformed Lagrangian and symmetry transformations
can be written more compactly in terms of auxiliary
covariant derivatives (\ref{cov}) where now $\Delta$ is
given by (\ref{Delta2}). The deformed Lagrangian
for two vector multiplets and the double tensor multiplets
reads then
\beq
L=\9 L0_{TT}-\sfrac 14 \7F^A_{\mu\nu}\7F^{\mu\nu A}
+\sfrac 12 \7D_\mu\phi^A\7D^\mu \5\phi^A
-2\Ii \lambda^{iA}\sigma^\mu \7D_\mu \5\lambda^{iA}
\label{L2}
\eeq
with $\9 L0_{TT}$ as in (\ref{TTfree}) and 
\beq
\7F^A_{\mu\nu}=\7D_\mu A_\nu^A-\7D_\nu A_\mu^A
=F^A_{\mu\nu}+g^i h_\mu^i\ep^{AB}A_\nu^B-g^i h_\nu^i\ep^{AB}A_\mu^B\ .
\eeq
The deformed gauge transformations are
\bea
&&\delta_\vep A_\mu^A=\6_\mu \vep^A+g^i h_\mu^i\ep^{AB}\vep^B
=:\7D_\mu \vep^A
\nonumber\\
&&\delta_\vep B_{\mu\nu}^i = 
\sfrac 14 g^i\vep^A \ep^{AB}\ep_{\mu\nu\rho\sigma}\7F^{\rho\sigma B}
+\6_\mu \vep^i_\nu-\6_\nu \vep^i_\mu
\nonumber\\
&&\delta_\vep(\mbox{other fields})=0
\eea
and the deformed supersymmetry transformations are
\bea
D_\alpha{}^i A^A_\mu &=& \ep^{ij}(\sigma_\mu \5\lambda^{jA})_\alpha
- \ep^{AB} \Gamma^i_\alpha A^B_\mu
\nonumber\\
D_\alpha{}^i \phi^A &=& 2 \lambda^{iA}_\alpha
- \ep^{AB} \Gamma^i_\alpha \phi^B
\nonumber\\
D_\alpha{}^i \5\phi^A &=& - \ep^{AB} \Gamma^i_\alpha \5\phi^B
\nonumber\\
D_\alpha{}^i \lambda^{jA}_\beta &=& -\sfrac{\Ii}2\, \ep^{ij}
                    \sigma^{\mu\nu}_{\alpha\beta}\7F^A_{\mu\nu}
- \ep^{AB} \Gamma^i_\alpha \lambda^{jB}_\beta
\nonumber\\
D_\alpha{}^i \5\lambda^{jA}_\da &=& -\sfrac {\Ii}2\, \delta^{ij}\, 
\7D_{\alpha\da}\5\phi^A- \ep^{AB} \Gamma^i_\alpha \5\lambda^{jB}_\da
\nonumber\\
D_\alpha{}^i B_{\mu\nu}^j&=&
(\ep^{ij}\sigma_{\mu\nu}\chi+\delta^{ij}\sigma_{\mu\nu}\psi
+\Ii g^j\ep^{AB}\5\phi^A\sigma_{\mu\nu}\lambda^{iB}
+\Ii g^j\ep^{AB}\ep^{ik}A_{[\mu}^A\sigma^{}_{\nu]}\5\lambda^{kB})_\alpha
\nonumber\\
D_\alpha^i &=&\9 D0_\alpha{}^i\ \mbox{on the other fields.}
\eea
Again, one may check that (\ref{algebra}) holds.
Elimination of the auxiliary fields $h_\mu^i$ is more cumbersome
than in the first model but it is clearly possible.

\subsection{Third example (type A)}

In the formulation with the auxiliary fields, (\ref{L300}) becomes
\beq
\9 L1=
\sfrac 12 g^k \ep^{ij} h_\mu^i h_\nu^j
B_{\rho\sigma}^k\ep^{\mu\nu\rho\sigma}
+2g^i h_\mu^i\ep^{jk} a^j\6^\mu a^k 
+\Ii g^i h_\mu^i
(\chi\sigma^\mu\5\psi-\psi\sigma^\mu\5\chi),
\quad g^i\in\mathbb{R}
\label{L1a}
\eeq
where we have renamed $k^i_1$ to $g^i$.
This is again of the form (\ref{d1}), with
\[
j^{\mu i}=g^i j^\mu,\quad
j^\mu=
\ep^{kj}h_\nu^jB_{\rho\sigma}^k\ep^{\mu\nu\rho\sigma}
+2\ep^{jk} a^j\6^\mu a^k
+\Ii\chi\sigma^\mu\5\psi 
-\Ii\psi\sigma^\mu\5\chi\ .
\]
This time one gets
\[
\9 \delta0_\vep \9 L1\simeq -\sum_\Phi (\9 \delta1_\vep\Phi)
\,\frac{\delta \9 L0}{\delta \Phi}\ ,\quad
\9 D0_\alpha{}^i\9 L1\simeq -\sum_\Phi
(\9 D1_\alpha{}^i\Phi)\,\frac{\delta \9 L0}{\delta \Phi}
\]
where
\bea
\9 \delta1_\vep B_{\mu\nu}^i &=& 
\ep^{ij}g^k (h_\mu^j \vep^k_\nu -h_\nu^j \vep^k_\mu),
\quad\9 \delta1_\vep(\mbox{other fields})=0
\label{d9}\\[4pt]
\9 D1_\alpha{}^i a^j&=&
\Gamma^i_\alpha \ep^{jk} a^k
\nonumber\\
\9 D1_\alpha{}^i B_{\mu\nu}^j&=&
\Ii [g^j\sigma_{\mu\nu}(\ep^{ik}\chi-\delta^{ik}\psi)a^k
+\sfrac12 g^k(\ep^{ij}\psi-\delta^{ij}\chi)B_{\mu\nu}^k
]_\alpha
\nonumber\\
\9 D1_\alpha{}^i h_\mu^j&=&
\sfrac{\Ii}2 g^j (\delta^{ik}\chi-\ep^{ik}\psi)h_\mu^k
\nonumber\\
\9 D1_\alpha{}^i\chi_\beta&=&\Gamma^i_\alpha \psi_\beta
\nonumber\\
\9 D1_\alpha{}^i\psi_\beta&=&-\Gamma^i_\alpha \chi_\beta
\nonumber\\
\9 D1_\alpha{}^i\5\chi_\da&=&\Ii g^j h_{\alpha\da}^j \ep^{ik}a^k
+\Gamma^i_\alpha \5\psi_\da
\nonumber\\
\9 D1_\alpha{}^i\5\psi_\da&=&-\Ii g^j h_{\alpha\da}^j a^i
-\Gamma^i_\alpha \5\chi_\da
\label{d10}\eea
with $\Gamma^i_\alpha$ as in (\ref{Gamma}).
In contrast to the first two examples, $\9 D1_\alpha{}^i h_\mu^i$ does
not vanish. This makes this example more complicated. 
To determine $\9 L2$, one must compute $\9 \delta1_\vep \9 L1$
and $\9 D1_\alpha{}^i \9 L1$. The results are
\bea
\9 \delta1_\vep \9 L1\simeq 0,\quad
\9 D1_\alpha{}^i \9 L1\sim-\9 D0_\alpha{}^i\9 L2
\label{d11}
\eea
where
\bea
\9 L2&=&g^ig^j  h^{\mu i} h_\mu^j a^ka^k-
g^ig^i a^jh^{\mu j} a^k h_\mu^k+
\sfrac 13 g^ig^i \ep^{jk} a^j \6_\mu a^k\ep^{lm}a^l\6^\mu a^m
\nonumber\\
&&
-g^ig^i a^j h_\mu^j (\psi\sigma^\mu\5\chi+\chi\sigma^\mu\5\psi)
+\sfrac {\Ii}2 g^ig^i \ep^{jk} a^j \6_\mu a^k
(\chi\sigma^\mu\5\psi-\psi\sigma^\mu\5\chi)
\nonumber\\
&&
-\sfrac 14 g^ig^i (\psi\psi\5\chi\5\chi+\chi\chi\5\psi\5\psi).
\label{d12}
\eea
Note that $\9 L2$ is invariant under $\9 \delta0_\vep$.
Hence the two equations
(\ref{d11}) are compatible. (\ref{d11}) shows also
that there is no 
second order deformation of the gauge transformations.
However, the second order deformation 
of the supersymmetry transformations does not vanish in this case
because the free field equations
are used in the second equation (\ref{d11}) (recall that 
$\sim$ stands for on-shell equality in the free theory modulo
total derivatives). To go on along the above lines, one must
determine $\9 D2_\alpha{}^i$, compute $\9 D2_\alpha{}^i \9 L1
+\9 D1_\alpha{}^i \9 L2$ etc.

Of course, it is at this stage not completely clear
whether the deformation exists at all orders $\geq 3$.
However, there are good reasons to assume that it does exist.
A simple inductive argument shows that 
all $\9 Lr$ which one would get by
continuing the above procedure have 
dimension 4 when one assigns dimension
$-1$ to the coupling constants $g^i$. Hence, $\9 Lr$ would be
a linear combination of field monomials $\9 Mr$ of dimension $r+4$, 
with coefficients of order $r$ in the $g^i$.
Furthermore $\9 Mr$ would have degree $r+2$ in the fields.
Note that $\9 L2$ involves only
the fields $a^i,h_\mu^i,\psi,\chi,\5\psi,\5\chi$ (but not
the $B_{\mu\nu}^i$). Assume now that $\9 L3$, $\9 L4$, \dots
can be chosen such that they involve also only these fields. The field
monomials $\9 Mr$, $r\geq 2$ would then fulfill
\[
r\geq 2:\quad N_h+N_a+N_f=r+2,\quad N_\6+2N_h+N_a+\sfrac 32 N_f=r+4
\]
where $N_h$, $N_a$, $N_f$ are the degrees in the $h_\mu^i$, $a^i$
and the fermions respectively, and $N_\6$ is the number of derivatives.
This implies
\[
r\geq 2:\quad N_\6+N_h+\sfrac 12 N_f=2.
\]
Hence, each $\9 Mk$ could only have $(N_\6,N_h,N_f,N_a)=
(2,0,0,k+2)$, $(1,1,0,k+1)$, $(1,0,2,k)$, $(0,2,0,k)$,
$(0,1,2,k-1)$ or $(0,0,4,k-2)$. Modulo trivial
terms, this would yield
\bea
r\geq 2:\quad
\9 Lr&=&\9 Ar{}^{ij}(g,a)\6_\mu a^i \6^\mu a^j 
+\9 Br{}^{ij}(g,a)h_\mu^i \6^\mu a^j 
+ \9 Cr{}^{ij}(g,a) h_\mu^i h^{\mu j}
\nonumber\\
&&
+\9 Dr{}^{ijk}(g,a) f^i\sigma^\mu\5f^j \6_\mu a^k
+\9 Er{}^{ijk}(g,a)  f^i\sigma^\mu\5f^j h_\mu^k
\nonumber\\
&&
+\9 Fr{}^{ijkl}(g,a) f^if^j\5f^k\5f^l
\label{r>1}
\eea
where $\{f^i\}=\{\psi,\chi\}$ and $\9 Ar{}^{ij}(g,a),\dots ,
\9 Fr{}^{ijkl}(g,a)$ are polynomials in the $g^i$ and the undifferentiated
$a^i$. In particular the $h_\mu^i$ could be 
eliminated algebraically also in the deformed theory
and the complete deformation of the gauge transformations 
would be given by $\delta_\vep=\9 \delta0_\vep+\9 \delta1_\vep$.
The supersymmetry transformations may of course receive
higher order contributions $\9 Dr_\alpha{}^i$.
One can check that the supersymmetry algebra takes to first order
again the standard form (\ref{algebra}). 

{\em Remark.}
A sufficient condition for the existence of $\9 Lr$
as in (\ref{r>1}) to all orders
can be formulated in cohomological terms.
As remarked at the end of section \ref{DEF},
the deformation problem can be reformulated as
a cohomological problem for an extended BRST differential $\9 s0$
which encodes the zeroth order gauge and supersymmetry
transformations.
It is then easy to show that
the existence of $\9 Lr$ as in (\ref{r>1}) is controlled
by the on-shell cohomology of $\9 s0$ (modulo
total derivatives)
in a space of Poincar\'e invariant local field
polynomials in the $a^i,h_\mu^i,\psi,\chi,\5\psi,\5\chi$
and their derivatives which depend in addition
linearly on the supersymmetry ghosts (as it is
the cohomology at ghost number 1 which enters here;
the ghosts of
the gauge transformations do not come into play here
because $\9 \delta0_\vep$ and $\9 \delta1_\vep$ act
nontrivially only on the $B_{\mu\nu}^i$).
Vanishing of that cohomology would guarantee the
existence of $\9 Lr$ as in (\ref{r>1}) for all $r$.
It is in fact reasonable to believe that this cohomology indeed
vanishes because its counterpart for N=1 supersymmetric
models with linear multiplets vanishes (this can be shown
as in \cite{susycohom}, owing to the fact that free linear
multiplets have ``QDS-structure'' on-shell).

\mysection{Comment on an N=1 superfield construction}\label{superfields}

It has been already mentioned that the free N=2 double tensor 
multiplet consists of two free N=1 linear multiplets. Similarly,
a free N=2 vector multiplet consists of one free N=1 vector multiplet
and one free N=1 chiral multiplet, while a free N=2  hypermultiplet
consists of two N=1 chiral multiplets. It is well-known that
all these N=1 multiplets have off-shell superfield descriptions involving 
auxiliary fields. These superfields might also be used to construct 
N=2 supersymmetric interactions involving the double tensor multiplet.
A promising step into that direction is the construction in \cite{BT}.
It provides N=1 supersymmetric interactions of the sought type
between N=1 linear multiplets and N=1 vector multiplets, and
it can be extended so as to include
N=1 chiral multiplets. 
Some of the resulting models will even be N=2 supersymmetric --
an example has been
recently given in \cite{T}. However, it is far from
obvious how one can sieve out
the models with a second supersymmetry systematically.

\paragraph{Acknowledgements.} The author acknowledges discussions with
Sergei Kuzenko and Ulrich Theis and was supported by
a DFG habilitation grant.

\end{document}